# Title: Tunable exciton valley-pseudospin orders in moiré Bose-Hubbard model

**Authors:** Richen Xiong[1], Samuel L. Brantly[1], Kaixiang Su[1], Jacob H. Nie[1], Zihan Zhang[1], Rounak Banerjee[2], Hayley Ruddick[2], Kenji Watanabe[3], Takashi Taniguchi[4], Sefaattin Tongay[2], Cenke Xu[1], Chenhao Jin[1]*

**Affiliations:**

[1]*Department of Physics, University of California at Santa Barbara, Santa Barbara, CA, USA*

[2]*School for Engineering of Matter, Transport, and Energy, Arizona State University, Tempe, AZ, USA.*

[3]*Research Center for Functional Materials, National Institute for Materials Science, Tsukuba, Japan.*

[4]*International Center for Materials Nanoarchitectonics, National Institute for Materials Science, Tsukuba, Japan*

\* Corresponding author. Email: jinchenhao@ucsb.edu

**Abstract:** Spin and charge are the two most important degrees of freedom of electrons. Their interplay lies at the heart of numerous strongly correlated phenomena including Hubbard model physics and high temperature superconductivity. Such interplay for bosons, on the other hand, is largely unexplored in condensed matter systems. Here we demonstrate a unique realization of the spin-½ Bose-Hubbard model through excitons in a semiconducting moiré superlattice. We find evidence of a transient in-plane ferromagnetic (FM-*xy*) order of exciton "spin" – here valley pseudospin – around exciton filling $v_{ex} = 1$, which transitions into a FM-*z* order both with increasing exciton filling and a small magnetic field of ~10 mT. The phase diagram is different from the fermion case and is qualitatively captured by a simple phenomenological model, highlighting the unique consequence of Bose-Einstein statistics. Our study paves the way for engineering exotic phases of matter from spinor bosons, as well as for unconventional devices in optics and quantum information science.

The interplay between spin and charge degrees of freedom in strongly correlated systems gives rise to a plethora of exotic phenomena. A notable example is the Hubbard model, where the competition between spin super-exchange interaction and charge hopping leads to a rich phase diagram that may account for the emergence of high temperature superconductivity from a doped antiferromagnetic (AFM) correlated insulator[1,2]. Moiré superlattices recently emerged as a powerful playground to engineer correlated phenomena[3]. Along with the strong light-matter interaction and unique optical selection rules[4,5], correlated excitons in semiconducting moiré systems hold promises for novel applications in photonics and valleytronics[6–8]. However, while various magnetic orderings and correlated phases of electrons are reported, such as correlated insulator[9–13], superconductivity[14–16], intrinsic[17,18] or exciton-mediated ferromagnetism[19], and fractional quantum anomalous Hall states[20–23]; correlated phases of excitons remain unexplored until very recently[24–27], and exciton "magnets" with "spin" orders have not been demonstrated.

Here we observe an intriguing phase diagram of interlayer-exciton "spin" orders in $WSe_2/WS_2$ moiré superlattices near one exciton per lattice site. Spin-up and spin-down exciton "spins" correspond to (and hereafter refer to) the *K* and *K'* valleys – two degenerate but inequivalent corners of the hexagonal Brillouin zone – and are related by time-reversal symmetry[4]. Similar to magnetism of real spins, exchange interaction between excitons of different valley pseudospins can lead to spontaneous ordering of the valley degree of freedom. For example, an out-of-plane ferromagnetic (FM-*z*) spin order polarizes spins to the same out-of-plane direction, which, in the exciton context, corresponds to a state where excitons are spontaneously polarized to the same valley; On the other hand, an in-plane spin is a coherent superposition of spin-up and spin-down. Therefore, excitons in an in-plane (*xy*) order are each a superposition between the two valleys of equal amplitude[28,29]. To probe exciton "spin" order, we use a pump probe spectroscopy[24] that isolates the low energy excitations of the system (Fig. 1a, see Methods: pump probe spectroscopy). Similar to electrical capacitance measurements[30,31], the DC pump light controls the background exciton density and maintains the quasi-equilibrium state, while the AC probe light injects a small perturbation of extra excitons and isolates their responses through lock-in detection. Owing to the optical selection rules in transition metal dichalcogenides, left- and right-circularly polarized (LCP and RCP) light selectively couple to spin-up and spin-down excitons[4,32,33]. This allows us to independently control spin of the background excitons and the extra injected excitons through polarization of the pump and probe light, respectively; and obtain spin-resolved system response from polarization-resolved photoluminescence (PL) detection. We can thereby directly create and probe low energy spin excitations of the system.

**Spin-½ Bose-Hubbard model**

Figure 1b shows the pump probe PL spectrum of a 0-degree-aligned $WSe_2/WS_2$ moiré device D1 with unpolarized pump, probe light and PL detection (the electron density is kept at 0 throughout this study). Unpolarized light couples to the total population of excitons[4,32,34]. The measurement therefore directly obtains the energy to remove one exciton from the system, i.e., its chemical potential. A sudden jump of exciton chemical potential is observed at one exciton per moiré site ($v_{ex} = 1$, $n_{ex} = 1.9 \times 10^{12}$ cm$^{-2}$, see Methods: Calibration of background exciton density), corresponding to a bosonic correlated insulator state[24]. The low and high energy peaks, labelled peak I and II, correspond to PL emissions from a singly occupied site and a doublon site (site with two excitons). Their energy shift of ~30 meV provides a direct measurement of exciton-exciton on-site repulsion and indicates the strong correlation between excitons[24].

We then switch experimental configurations to inspect spin excitations of excitons. The minimum model to account for exciton "spins" is a two-component Bose Hubbard model[35], given by

$$H = \sum_{<i,j>,\alpha} -t b^{\dagger}_{i,\alpha} b_{j,\alpha} + h.c. + \sum_{i,\alpha} U(n_{i,\alpha} - 1/2)^2 + \sum_i V n_{i,1} n_{i,2}$$

Here the $\alpha=1,2$ label $K$ and $K'$ valley pseudo-spin of excitons, $t$ is hopping between nearest neighboring sites, and the interactions between excitons consist of intra-species repulsion $U$ and inter-species repulsion $V$. To establish such a model and separately determine $U$ and $V$, we use linearly polarized pump light to generate equal population of two "spins" in the background, an LCP probe light to selectively inject extra $K$ valley ("spin"-up) excitons, and monitor "spin"-resolved responses by separately collecting RCP ($K'$ valley) and LCP ($K$ valley) PL from the probe light only (Fig. 1, c and d). The $K'$ and $K$ responses are rather similar at $v_{ex}<1$ (Fig. 1e), which can be understood in the single exciton picture from a short valley lifetime that quickly relaxes valley polarization. We directly capture such relaxation process by time-resolved pump probe PL measurements (Fig. 1g). The pulsed probe light selectively injects $K$ excitons at time zero, and the $K'$ response remains unchanged. Afterwards the valley polarization quickly disappears over time, resulting in similar overall responses from the two valleys (Fig. 1e).

In contrast, the two valleys' responses become dramatically different above $v_{ex} = 1$ (Fig. 1f). Most strikingly, their responses have opposite signs for peak I. The negative $K'$ response indicates that adding extra $K$ excitons will decrease the number of singly occupied $K'$ sites. Such behavior is incompatible with the single exciton picture where adding $K$ excitons always increase both $K$ and $K'$ exciton populations[34,36] and is instead a unique consequence of exciton correlation. Our observation can be naturally understood from the Bose Hubbard model: the $K$ excitons injected by the probe form doublons at $v_{ex} > 1$, which will decrease the number of singly occupied sites by converting them into doublon sites. The decrease in $K'$ sites therefore indicates that $K$ excitons selectively form doublon sites with $K'$ excitons (Extended Data Fig. 2a). This is further confirmed by the perfect valley balance in doublon emission (peak II) regardless of experimental configuration (Extended Data Fig. 2b), which requires $K$ and $K'$ excitons to be symmetric within any doublons.

Our results thus unambiguously establish a "spin"-dependent on-site repulsion between excitons. The ~30 meV jump of exciton chemical potential at $v_{ex} = 1$ (Fig. 1b) corresponds to the opposite-"spin" repulsion $V$, while the same-"spin" repulsion $U$ is much greater than $V$. Consequently, doublons only form by two excitons of opposite "spins" like electrons in a Fermi-Hubbard model, which offers a rare realization of spin-½ Bose-Hubbard model.

**Inter-site spin-dependent interactions between excitons**

Next, we investigate inter-site spin interactions that may lead to spin orders. We vary the pump light polarization and keep the probe LCP. Different pump polarization maintains background excitons of different spins, while the LCP probe light always injects spin-up excitons. Any difference in the measured probe response can therefore directly reflect spin-dependent interaction between excitons. Fig. 2 a-c show the probe-induced spin imbalance (the difference between $K$ and $K'$ emission induced by the probe light) as a function of exciton fillings for LCP, RCP and linear pump respectively. See Extended Data Fig. 4 for results on another device D2. Peak II always shows zero spin-imbalance signal, as expected for doublon emission (Extended Data Fig. 2). Peak I is insensitive to pump polarization at low exciton density, suggesting negligible spin interaction effects. At increasing exciton density, in contrast, the signals vary dramatically with pump polarization. Under both LCP and RCP pump (Fig. 2, a and b), we observe a sharp signal enhancement at $v_{ex} \sim 1.1$ followed by a quick drop at $v_{ex} > 1.2$ (black arrows). Both features are absent in the linear pump case (Fig. 2c), indicating their origin from spin interactions. While the strongest spin interaction in the system is the on-site AFM exchange, it cannot account for the symmetric behaviors between the LCP and RCP pump or the sensitive filling dependence (see Methods: Spin-1/2 Bose Hubbard model). These features therefore indicate rapidly changing inter-site spin interactions when doping slightly away from a bosonic correlated insulator.

We first investigate the feature at $v_{ex} = 1.1$ and monitor its evolution under an out-of-plane magnetic field $B_z$. Fig. 2 d-f shows the magnetic field-dependent spin imbalance spectra for fixed $v_{ex} = 1.1$ and different pump polarization. Surprisingly, the probe-induced spin imbalance under CP pumps or linear pump are either suppressed or enhanced by an order of magnitude, respectively, upon applying a tiny magnetic field of 5 mT. Such sensitive magnetic field dependence and low saturation field (~20 mT) of exciton spin polarization have not been reported before[37–40], and generally indicates adjacent phase transitions with strong spin fluctuations[41].

To further confirm the phase transition, we also perform pump-only PL measurement. Such measurement collects PL from all background excitons in the system and is therefore less sensitive to spin interactions than the pump-probe measurement. However, a spin order will affect not only low energy excitations but all excitons in the system and should therefore be observable in such measurements. Fig. 3a shows example spectra of $K$ and $K'$ PL at $v_{ex} = 1.39$ with RCP pump, from which we obtain the PL raw helicity $\eta_{PL} = \frac{I_{K,PL} - I_{K',PL}}{I_{K,PL} + I_{K',PL}}$. $I_{K,PL}$ and $I_{K',PL}$ are the PL emission intensity from $K$ and $K'$ excitons (peak I in Fig. 3a), which are proportional to the number of singly occupied $K$ and $K'$ sites, respectively. Fig. 3b summarizes $\eta_{PL}$ under different pump polarization and exciton fillings. To reveal spin orders, we introduce generalized helicity GH = $\eta_{PL}/\eta_{pump}$, where $\eta_{pump}$ is the helicity of pump light (see Methods: Data analysis and Extended Data Fig. 9). In the case of no spin order, the LCP and RCP components of pump light should independently contribute to PL emission, and therefore GH will be a constant over pump polarization. A spin order, on the other hand, generates a mean field that depends on all exciton spins in the system and thus on the pump polarization. Consequently, GH will change with pump helicity $\eta_{pump}$.

Figure 3, c and d show GH and normalized GH over pump polarization (helicity) at different exciton fillings. Since GH is not well-defined at $\eta_{pump}=0$ (linear pump), only data at $|\eta_{pump}|>0.02$ are obtained experimentally (symbols); and GH at $\eta_{pump}=0$ can be extrapolated from the limit of $\eta_{pump}\to 0$ (see Methods: Data analysis). At low exciton density GH is indeed a constant over pump polarization. In contrast, GH becomes a "Λ" shape at $v_{ex} = 1.1$ and quickly transitions into a "V" shape at $v_{ex} > 1.25$, echoing the two features in the pump probe PL spectra (Fig. 2, a and b). When we further apply an out-of-plane magnetic field, the constant GH at low exciton filling remains intact (Fig. 3e). The "Λ" shape GH at $v_{ex} = 1.1$, on the other hand, changes dramatically and becomes a "V" shape under both ±30 mT field (see Extended Data Fig. 7a for data at 30 mT). Such sensitive and symmetric magnetic field dependence is consistent with the pump-probe results and again signifies an adjacent phase transition. We have also measured normalized GH at 60 K as a reference (Fig. 3f), which is flat over the whole exciton density range. This further confirms the origin of nontrivial shapes in GH from exciton spin orders.

**Tunable transient exciton spin orders**

To unravel the nature of the exciton spin orders, we performed detailed magnetic field dependence at $v_{ex} = 1.1$ and 1.3. Figure 4a and 4b show the PL raw helicity and GH at $v_{ex} = 1.1$ from 0 to 30 mT. Intriguingly, the GH at $B_z >20$ mT exceeds unity in a wide pump polarization range. A GH > 1 means that the spin polarization of the system is higher than the pump. This cannot be explained by field-induced symmetry breaking between the two spins, which would favor one spin over the other and lead to asymmetric GH between RCP and LCP pump. In contrast, the observed GH is symmetric against $\eta_{pump}=0$ and exceeds unity on both sides (Fig. 4b). If excitons in the system were not decaying over time – equivalently, all PL emissions are re-absorbed by the system – the system would keep amplifying the spin polarization. A tiny initial spin-up/down injection would then eventually develop into a close to fully spin-up/down state. Such spontaneous spin polarization is the hallmark of an FM-z order. On the other hand, because here excitons are in a quasi-equilibrium between decaying and pumping, any system memory is lost over the exciton lifetime and there should not be hysteresis. Hence all orders identified experimentally are of transient nature at the timescale of exciton decay.

Our observation indicates a transient FM-z order of excitons at $v_{ex} = 1.1$ and magnetic field $B_z >20$ mT. The zero-field state at $v_{ex} = 1.1$ is more exotic. Phenomenologically, it shows opposite behaviors from the high field FM-z order in both pump probe and GH measurements (Fig. 2a-c and Fig. 4b), indicating distinct exciton spin states. Its extremely sensitive magnetic field dependence and low saturation field (~20 mT) is particularly surprising. The most well-known effect from an out-of-plane magnetic field is the Zeeman splitting that lifts the degeneracy between the two exciton spins. Similarly, the dominant effect of magnetic field on excitons is an energy splitting between the two valleys, termed "valley Zeeman effect"[37,38], with a g-factor of 4 in monolayer TMD. However, the splitting should be <~0.1 meV under 10 mT[34], which is too small compared to the expected energy scale of spin interaction (~1 meV, see Methods: Discussions on magnetic field dependence). In addition, applying positive and negative $B_z$ should lead to opposite Zeeman splitting. Instead, in our experiment they have mostly symmetric effects and eventually result in a transition into the same FM-$z$ order (Fig. 2 d-f, Fig. 3e and Extended Data Fig. 7a). Both observations exclude a simple linear coupling between the Zeeman field and the order parameter and suggest a finite in-plane component in the zero-field order, i.e., an $xy$ order (For more discussions, see Methods: $xy$ order).

Indeed, such phase transition can fully explain our experimental observations. Under linear pump, the $xy$ spin order creates an in-plane mean field, which will efficiently mix up and down spins and suppress spin polarization. On the other hand, spins under CP pumps are initialized to be along the $z$ direction and the in-plane mean field is weaker. We therefore expect a stronger suppression of spin polarization near linear pump and a weaker suppression near CP pumps, i.e., a "Λ" shape GH (Fig. 4b). The high field FM-$z$ order, on the contrary, amplifies spin polarization. This leads to a sharp rise of $\eta_{PL}$ with $\eta_{pump}$ and a large GH>1 when $|\eta_{pump}|$ is small. At large $|\eta_{pump}|$, $\eta_{PL}$ saturates since it cannot exceed 1 (Fig. 4a); and GH is always smaller than 1 when $|\eta_{pump}|$ =1 (CP pumps). We therefore expect a "V" shape GH as observed experimentally (Fig. 4b). The transition region between the $xy$ and FM-$z$ orders at intermediate field is more complicated, where GH becomes asymmetric between positive and negative $\eta_{pump}$ or $B_z$ (Fig. 4b). This indicates extrinsic symmetry breaking between the two spins by the magnetic field and thus no well-defined order (see Methods: Discussions on magnetic field dependence).

We now turn to the feature at $v_{ex}$ >1.25. Its zero field behaviors are qualitatively similar to the high field behaviors at $v_{ex}$ = 1.1 in all measurements: pump probe measurement (Fig. 2) shows a stronger spin imbalance signal under linear pump compared to CP pump; GH shows a "V" shape with GH > 1 over a wide range of pump helicity. Upon applying an out-of-plane magnetic field, these behaviors are qualitatively unchanged and quantitatively enhanced. For example, GH is enhanced to a giant value of 2.3 near linear pump (Fig. 4d), corresponding to a rapid increase and saturation of spin polarization as $\eta_{pump}$ increases that can be clearly seen in the PL raw helicity (Fig. 4c). These results provide strong evidence that at $v_{ex}$ >1.25 the system is already in an FM-$z$ order without magnetic field, i.e., suggesting a filling-controlled transition between $xy$ to FM-$z$ order at $v_{ex}$ ~1.25.

The pump-probe measurement results (Fig. 2) are also naturally explained by the competition between the in-plane and out-of-plane spin interactions of excitons. At $v_{ex}$ =1.1, the dominant in-plane spin interactions under a linear pump rapidly quench out-of-plane spin imbalances, while a CP pump reduces such quench by forcing exciton spins to be out-of-plane. We therefore observe stronger probe-induced spin imbalance signals under both LCP and RCP pumps compared to the linear pump case. Upon increasing exciton filling and/or magnetic field $B_z$, the FM-$z$ spin interactions dominate and enhance spin imbalances under linear pump as the system is not fully polarized. Once all spins are polarized under CP pump, the system is not susceptible to spin excitations anymore and the probe-induced spin imbalance becomes vanishingly small.

We next measure the temperature dependence of these orders. Figure 5 a and b show normalized GH at $v_{ex}$ = 1.1 and 1.3, respectively, for temperatures from 3 to 60 K. The "Λ" shape GH at $v_{ex}$ = 1.1 and "V" shape GH at $v_{ex}$ = 1.3 disappear at around 35 and 50 K, respectively, indicating melting of the associated spin orders. To quantify the temperature dependence, we define $\Delta_{GH} = \frac{\text{GH(CP pump)} - \text{GH(linear pump)}}{\text{GH(CP pump)}}$. A positive and negative $\Delta_{GH}$ correspond to a "Λ" shape and "V" shape GH and indicates $xy$- and $z$- spin order, respectively. Fig. 5 c and d show the phase diagram of $\Delta_{GH}$ at 0 and -30 mT (see Extended Data Fig. 7b for 30 mT data). We also mark regions where GH (linear pump) >1 with dotted texture, which is the hallmark of an FM-$z$ order. At zero field, the system first enters an $xy$ order upon increasing exciton filling to $v_{ex}$ ~1 and then transitions into an FM-$z$ order at $v_{ex}$~1.25. At magnetic field of -30 mT the $xy$ order is suppressed, and the FM-$z$ order is favored over the filling range of $v_{ex}$>1.1. Its melting temperature keeps increasing with the exciton filling.

**Discussions and outlook**

Our observations provide strong evidence of phase transitions from a (transient) $xy$ order to FM-$z$ order driven by both exciton filling and magnetic field. This can be naturally understood in a spin-½ Bose-Hubbard model from competitions between the super-exchange effect and Nagaoka-type kinetic ferromagnetism[18,42]. Our pump-probe measurements establish WSe$_2$/WS$_2$ moiré superlattice as a spin-½ Bose Hubbard model, which has been predicted to host a ground state of FM-$xy$ order at $v_{ex}$ = 1 as the virtual hopping of bosons gives rise to an FM in-plane super-exchange interaction $J_\perp$[35,43]. Upon further doping, the kinetic energy of extra bosons would favor an FM-$z$ order, similar to Nagaoka ferromagnetism in Fermi Hubbard model[18,42,44]. We reveal the essential physics near $v_{ex}$=1 using a phenomenological spin-½ XXZ model on a triangular lattice, where the effect of adding excitons to the $v_{ex}$=1 correlated insulator is captured by a $z$-exchange interaction $J_z$ that increases with doping (see Methods: Theoretical phase diagram and Supplementary Materials). Figure 5e shows the phase diagram predicted by this phenomenological model, which matches well with the experimental one. A transition from FM-$xy$ to FM-$z$ order is expected with increasing doping (and $J_z$). In addition, the system is very sensitive to a Zeeman field $B_z$ near the transition; and a weak $B_z$ would favor the FM-$z$ over the FM-$xy$ order. Intriguingly, both the FM-$xy$ and FM-$z$ orders are unique consequences of Bose-Einstein statistics – it is well established that the super-exchange interaction in a Fermi Hubbard model is antiferromagnetic (AFM) along all directions[1,44], and the Nagaoka FM is also isotropic instead of favoring the $z$ direction[42,44].

Our study establishes semiconducting moiré superlattices as an intriguing platform to realize exotic states of excitons, which will open up novel device concepts in photonics and quantum information science. For example, an FM-$z$ order not only stores but also amplifies valley polarization, which may serve as a cornerstone of memory and error correction code[45,46]. The extremely sensitive magnetic field and exciton filling dependence are consequences of phase transitions and go beyond the single particle limit, which may enable efficient light source control and optical gates similar to phase-change transistors in electronics[47,48]. In addition, the two-component Bose-Hubbard model can potentially support a plethora of much more exotic phases beyond the ferromagnetic orders. For instance, it was shown numerically that a supersolid can be realized in the effective XXZ spin-1/2 model with one boson per-site[43]. Away from the vicinity of one boson per site, the two flavors of hard-core bosons can be mapped to an SU(4) spin model (with anisotropies, see Supplementary Materials), which is a system that has attracted enormous interest in the past few decades and hosts various intriguing phases[49–55].

**Reference:**

1. Lee, P. A., Nagaosa, N. & Wen, X. G. Doping a Mott insulator: Physics of high-temperature superconductivity. *Rev Mod Phys* **78**, 17–85 (2006).
2. Scalapino, D. J. A common thread: The pairing interaction for unconventional superconductors. *Rev Mod Phys* **84**, 1383–1417 (2012).
3. Kennes, D. M. *et al.* Moiré heterostructures as a condensed-matter quantum simulator. *Nat Phys* **17**, 155–163 (2021).
4. Xu, X., Yao, W., Xiao, D. & Heinz, T. F. Spin and pseudospins in layered transition metal dichalcogenides. *Nat Phys* **10**, 343–350 (2014).
5. Mak, K. F., Xiao, D. & Shan, J. Light–valley interactions in 2D semiconductors. *Nat Photonics* **12**, 451–460 (2018).
6. Regan, E. C. *et al.* Emerging exciton physics in transition metal dichalcogenide heterobilayers. *Nat Rev Mater* **7**, 778–795 (2022).
7. Ciarrocchi, A., Tagarelli, F., Avsar, A. & Kis, A. Excitonic devices with van der Waals heterostructures: valleytronics meets twistronics. *Nat Rev Mater* **7**, 449–464 (2022).
8. Mak, K. F. & Shan, J. Photonics and optoelectronics of 2D semiconductor transition metal dichalcogenides. *Nat Photonics* **10**, 216–226 (2016).
9. Cao, Y. *et al.* Correlated insulator behaviour at half-filling in magic-angle graphene superlattices. *Nature* **556**, 80–84 (2018).
10. Tang, Y. *et al.* Simulation of Hubbard model physics in WSe2/WS2 moiré superlattices. *Nature* **579**, 353–358 (2020).
11. Regan, E. C. *et al.* Mott and generalized Wigner crystal states in WSe2/WS2 moiré superlattices. *Nature* **579**, 359–363 (2020).
12. Shimazaki, Y. *et al.* Strongly correlated electrons and hybrid excitons in a moiré heterostructure. *Nature* **580**, 472–477 (2020).
13. Jin, C. *et al.* Stripe phases in WSe2/WS2 moiré superlattices. *Nat Mater* **20**, 940–944 (2021).
14. Cao, Y. *et al.* Unconventional superconductivity in magic-angle graphene superlattices. *Nature* **556**, 43–50 (2018).
15. Oh, M. *et al.* Evidence for unconventional superconductivity in twisted bilayer graphene. *Nature* **606**, 494–500 (2021).
16. Kim, H. *et al.* Evidence for unconventional superconductivity in twisted trilayer graphene. *Nature*

**606**, 494–500 (2022).

17. Anderson, E. *et al.* Programming correlated magnetic states with gate-controlled moiré geometry. *Science (1979)* **381**, 325–330 (2023).

18. Ciorciaro, L. *et al.* Kinetic magnetism in triangular moiré materials. *Nature* **623**, 509–513 (2023).

19. Wang, X. *et al.* Light-induced ferromagnetism in moiré superlattices. *Nature* **604**, 468–473 (2022).

20. Cai, J. *et al.* Signatures of fractional quantum anomalous Hall states in twisted MoTe2. *Nature* **622**, 63–68 (2023).

21. Zeng, Y. *et al.* Thermodynamic evidence of fractional Chern insulator in moiré MoTe2. *Nature* **622**, 69–73 (2023).

22. Park, H. *et al.* Observation of fractionally quantized anomalous Hall effect. *Nature* **622**, 74–79 (2023).

23. Xu, F. *et al.* Observation of Integer and Fractional Quantum Anomalous Hall Effects in Twisted Bilayer MoTe2. *Phys Rev X* **13**, (2023).

24. Xiong, R. *et al.* Correlated insulator of excitons in WSe2/WS2 moiré superlattices. *Science (1979)* **380**, 860–864 (2023).

25. Park, H. *et al.* Dipole ladders with large Hubbard interaction in a moiré exciton lattice. *Nat Phys* **19**, 1286–1292 (2023).

26. Gao, B. *et al. Excitonic Mott Insulator in a Bose-Fermi-Hubbard System of Moiré WS2/WSe2 Heterobilayer*. https://arxiv.org/abs/2304.09731v1 (2023) doi:https://doi.org/10.48550/arXiv.2304.09731.

27. Lian, Z. *et al.* Valley-polarized excitonic Mott insulator in WS2/WSe2 moiré superlattice. *Nat Phys* **20**, 34–39 (2024).

28. Jones, A. M. *et al.* Optical generation of excitonic valley coherence in monolayer WSe 2. *Nat Nanotechnol* **8**, 634–638 (2013).

29. Ye, Z., Sun, D. & Heinz, T. F. Optical manipulation of valley pseudospin. *Nat Phys* **13**, 26–29 (2017).

30. Luryi, S. Quantum capacitance devices. *Appl Phys Lett* **52**, 501–503 (1988).

31. Eisenstein, J. P., Pfeiffer, L. N. & West, K. W. Compressibility of the two-dimensional electron gas: Measurements of the zero-field exchange energy and fractional quantum Hall gap. *Phys Rev B* **50**, 1760–1778 (1994).

32. Rivera, P. *et al.* Interlayer valley excitons in heterobilayers of transition metal dichalcogenides. *Nat Nanotechnol* **13**, 1004–1015 (2018).

33. Ciarrocchi, A., Tagarelli, F., Avsar, A. & Kis, A. Excitonic devices with van der Waals heterostructures: valleytronics meets twistronics. *Nat Rev Mater* **7**, 449–464 (2022).

34. Jiang, Y., Chen, S., Zheng, W., Zheng, B. & Pan, A. Interlayer exciton formation, relaxation, and transport in TMD van der Waals heterostructures. *Light Sci Appl* **10**, 72 (2021).

35. Altman, E., Hofstetter, W., Demler, E. & Lukin, M. D. Phase diagram of two-component bosons on an optical lattice. *New J Phys* **5**, 113–113 (2003).

36. Jin, C. *et al.* Ultrafast dynamics in van der Waals heterostructures. *Nat Nanotechnol* **13**, 994–1003 (2018).

37. Srivastava, A. *et al.* Valley Zeeman effect in elementary optical excitations of monolayer WSe2. *Nat Phys* **11**, 141–147 (2015).

38. Aivazian, G. *et al.* Magnetic control of valley pseudospin in monolayer WSe2. *Nat Phys* **11**, 148–152 (2015).

39. Nagler, P. *et al.* Giant magnetic splitting inducing near-unity valley polarization in van der Waals heterostructures. *Nat Commun* **8**, 1551 (2017).

40. Seyler, K. L. *et al.* Signatures of moiré-trapped valley excitons in MoSe2/WSe2 heterobilayers. *Nature* **567**, 66–70 (2019).
41. Sachdev, S. *Quantum Phase Transitions*. (Cambridge University Press, 2011). doi:10.1017/CBO9780511973765.
42. Nagaoka, Y. Ferromagnetism in a Narrow, Almost Half-Filled s Band. *Physical Review* **147**, 392–405 (1966).
43. Wang, F., Pollmann, F. & Vishwanath, A. Extended Supersolid Phase of Frustrated Hard-Core Bosons on a Triangular Lattice. *Phys Rev Lett* **102**, 017203 (2009).
44. Tasaki, H. The Hubbard model - an introduction and selected rigorous results. *Journal of Physics: Condensed Matter* **10**, 4353–4378 (1998).
45. Terhal, B. M. Quantum error correction for quantum memories. *Rev Mod Phys* **87**, 307–346 (2015).
46. Zhang, L. *et al.* Van der Waals heterostructure polaritons with moiré-induced nonlinearity. *Nature* **591**, 61–65 (2021).
47. Liu, K., Lee, S., Yang, S., Delaire, O. & Wu, J. Recent progresses on physics and applications of vanadium dioxide. *Materials Today* **21**, 875–896 (2018).
48. Sun, Z., Martinez, A. & Wang, F. Optical modulators with 2D layered materials. *Nat Photonics* **10**, 227–238 (2016).
49. Kugel', K. I. & Khomskiĭ, D. I. The Jahn-Teller effect and magnetism: transition metal compounds. *Soviet Physics Uspekhi* **25**, 231–256 (1982).
50. Wu, C., Hu, J. P. & Zhang, S. C. Exact SO(5) symmetry in the spin-3/2 fermionic system. *Phys Rev Lett* **91**, (2003).
51. Gorshkov, A. v. *et al.* Two-orbital SU(N) magnetism with ultracold alkaline-earth atoms. *Nat Phys* **6**, 289–295 (2010).
52. Xu, C. & Balents, L. Topological Superconductivity in Twisted Multilayer Graphene. *Phys Rev Lett* **121**, (2018).
53. Yuan, N. F. Q. & Fu, L. Model for the metal-insulator transition in graphene superlattices and beyond. *Phys Rev B* **98**, (2018).
54. Po, H. C., Zou, L., Vishwanath, A. & Senthil, T. Origin of Mott Insulating Behavior and Superconductivity in Twisted Bilayer Graphene. *Phys Rev X* **8**, (2018).
55. You, Y. Z. & Vishwanath, A. Superconductivity from valley fluctuations and approximate SO(4) symmetry in a weak coupling theory of twisted bilayer graphene. *NPJ Quantum Mater* **4**, (2019).
56. Wang, L. *et al.* One-dimensional electrical contact to a two-dimensional material. *Science (1979)* **342**, 614–617 (2013).
57. Mak, K. F. & Shan, J. Semiconductor moiré materials. *Nat Nanotechnol* **17**, 686–695 (2022).
58. Li, Y. Q., Ma, M., Shi, D. N. & Zhang, F. C. SU(4) Theory for Spin Systems with Orbital Degeneracy. *Phys Rev Lett* **81**, 3527–3530 (1998).
59. Wu, C. Competing Orders in One-Dimensional Spin- 3/2 Fermionic Systems. *Phys Rev Lett* **95**, 266404 (2005).
60. Penc, K., Mambrini, M., Fazekas, P. & Mila, F. Quantum phase transition in the SU(4) spin-orbital model on the triangular lattice. *Phys Rev B* **68**, 012408 (2003).
61. Xu, C. & Wu, C. Resonating plaquette phases in SU(4) Heisenberg antiferromagnet. *Phys Rev B* **77**, 134449 (2008).
62. Hermele, M., Gurarie, V. & Rey, A. M. Mott Insulators of Ultracold Fermionic Alkaline Earth Atoms: Underconstrained Magnetism and Chiral Spin Liquid. *Phys Rev Lett* **103**, 135301 (2009).

63. Corboz, P., Lajkó, M., Läuchli, A. M., Penc, K. & Mila, F. Spin-Orbital Quantum Liquid on the Honeycomb Lattice. *Phys Rev X* **2**, 041013 (2012).
64. Keselman, A., Bauer, B., Xu, C. & Jian, C.-M. Emergent Fermi Surface in a Triangular-Lattice SU(4) Quantum Antiferromagnet. *Phys Rev Lett* **125**, 117202 (2020).
65. Zhang, Y.-H., Sheng, D. N. & Vishwanath, A. SU(4) Chiral Spin Liquid, Exciton Supersolid, and Electric Detection in Moiré Bilayers. *Phys Rev Lett* **127**, 247701 (2021).
66. Mermin, N. D. & Wagner, H. Absence of Ferromagnetism or Antiferromagnetism in One- or Two-Dimensional Isotropic Heisenberg Models. *Phys Rev Lett* **17**, 1133–1136 (1966).

**Methods:**

**Device fabrication and characterization:** The dual-gated $WSe_2/WS_2$ devices were made by layer-by-layer dry transfer method[56]. Polarization-resolved second harmonic generation (SHG) was used to determine the crystalline angles between monolayer $WSe_2$ and $WS_2$ before stacking. hBN flakes with a thickness of around 20 nm were used as the gate dielectrics and few-layer graphite flakes were used as the gates and contact electrode. The whole stack was then released to a 90 nm $Si/SiO_2$ substrate with pre-patterned Au contacts. Extended Data Fig. 1a shows an optical image of 0-degree aligned $WSe_2/WS_2$ device D1. The twist angles were measured to be within $0 \pm 0.5°$, limited by experimental uncertainties. Extended Data Fig. 1 b and c show the gate-dependent absorption and PL characterization of the moiré bilayer at 3 K. At charge neutrality (~-0.05V), the PL features a single peak at 1.375eV from interlayer exciton emission, while the absorption shows three peaks from moiré intralayer excitons. At $v_e$=1 and $v_h$=1 (white arrows), the emission peak blueshifts suddenly and the absorption peaks show a kink, indicating the emergence of correlated insulator of charges. All measurements are performed at a base temperature of 3 K in 0°-aligned $WSe_2/WS_2$ device D1 unless specified.

**Pump probe spectroscopy:** The samples were mounted in a closed-cycle cryostat (Quantum Design, OptiCool). A continuous wave 660 nm diode laser was used as the pump light with beam size of around 100 $\mu m^2$. The large pump beam size ensures a homogeneous intensity in the center region that is inspected by the probe beam. A pulsed 680 nm light from a supercontinuum laser (YSL Photonics, 300 ps pulse duration, variable repetition rate) were used as the probe light. The beam size of probe light was around 4 $\mu m^2$. The probe intensity was kept below 30 $nW/\mu m^2$, while the pump intensity ranged from 0 to 3 $\mu W/\mu m^2$. To isolate the response from probe-created excitons, the probe light was modulated by an optical chopper at frequency of 10 Hz. The signal was detected by a liquid-nitrogen-cooled CCD camera coupled with a spectrometer (Princeton Instruments), which was externally triggered at 20 Hz and phase locked to the chopper. The spectra with and without the probe light were thereby obtained, and their difference gives the signals from the probe light only (see Fig. 1b for an example). To help isolate the probe light response, we also implement a spatial filter at a conjugate image plane of the sample, which only allows light from the probe-covered region to go through. For polarization-resolved measurements, the polarization of pump/probe/PL is controlled by broadband polarizers and half-wave plates. In time-resolved PL measurements (TRPL), the signals are collected by an avalanche photodiode (MPD PDM series) and analyzed by a time-correlated single photon counting module (ID Quantique ID1000). Since the pump light is CW while the probe light is pulsed, their contribution can be directly separated in the time domain and no AC modulation is needed. We thereby directly track the system's dynamical response to extra transient excitons at certain background exciton filling.

**Calibration of background exciton density:** We precisely calibrate the exciton density and filling at each pump intensity through time-resolved PL measurements. This is done in two steps. We first perform time-resolved PL (TRPL) measurement using the CW pump light as excitation light (Extended Data Fig. 10a). PL emission rate is a constant over time, as expected from CW excitation. This allows us to determine the emission rate at each pump intensity

([Extended Data Fig. 10c](#)). Next, we establish the relation between emission rate and exciton density by replacing the CW pump with a pulsed pump light (300 ps pulse duration, 1 MHz repetition rate) with the same wavelength (660nm) and beam profile. All other experimental configurations are also kept identical. [Extended Data Fig. 10b](#) shows the time-resolved PL using the pulsed excitation light of different pulse fluences. The decay dynamics changes with pulse fluence but is always much slower than the instrumental response function (IRF, [Extended Data Fig. 10b](#) inset). Therefore, the emission rate immediately after time-zero corresponds to the exciton density created by the pulsed excitation light without any relaxation, which can be directly obtained from the pulse fluence.

The above procedure allows us to reliably determine exciton density without complications from the exciton lifetime or relaxation dynamics. Since the system reaches quasi-equilibrium in a short time (<1ps), each measured emission rate uniquely corresponds to one exciton density at quasi-equilibrium, whether the excitation light is CW or pulsed. For example, at charge neutrality we identify $v_{ex}$ =1 at pump intensity of 0.406 μW/μm$^2$, which corresponds to an emission rate of 562 ([Extended Data Fig. 10c](#)). The same emission rate is achieved by the pulsed pump light with fluence $F$= 0.25 J/m$^2$ immediately after time zero ([Extended Data Fig. 10d](#)). The exciton density is directly obtained from the pulse fluence though $n_{ex} = \alpha F/(h\nu) = (1.9 \pm 0.2) \times 10^{12}$cm$^{-2}$, where $F$ is the pulse fluence, $\alpha=(0.023 \pm 0.002)$ is absorption of WSe$_2$ at 660nm using its dielectric function and considering the multi-layer structure of our device, $h\nu$ is the photon energy of 660nm light. $n_{ex}$ matches well with the expected exciton density $n_0 = \frac{2}{\sqrt{3}a_M^2}$ =1.9x10$^{12}$ cm$^{-2}$ at $v_{ex}$=1, where $a_M$ ~ 8 nm is the moiré periodicity considering 4% lattice mismatch and 0-degree twist angle. We calibrate the exciton density at all pump intensity and polarization following the above procedure.

**Data analysis**: To ensure the reliability of GH = $\eta_{PL}/\eta_{pump}$, we carefully calibrate the uncertainties in both $\eta_{pump}$ and $\eta_{PL}$. In our experiment, the pump light goes through a half-wave plate (HWP) and a quarter-wave plate (QWP) before impinging on the sample. The pump helicity $\eta_{pump}$ is controlled by the HWP angle $\theta$ and should ideally follow a simple sine function of $\eta_{pump} = \sin(\pi\frac{\theta-\theta_0}{45°})$. On the other hand, imperfect optics and/or alignment may result in deviation from such relation. To calibrate the uncertainty in $\eta_{pump}$, we directly measure the LCP and RCP components in the sample-reflected pump light using identical experimental configuration as measuring the LCP and RCP PL, as shown in [Extended Data Fig. 9a](#). This allows us to determine $\theta_0$ when the LCP and RCP components are equal. The extracted $\eta_{pump}$ ([Extended Data Fig. 9b](#)) shows a near-perfect match with the ideal sine relation (grey curve) and a relative standard deviation $\Delta\eta_{pump}/\eta_{pump}$ <2%.

To calibrate the uncertainty in $\eta_{PL}$, we measure $\eta_{PL}$ twice under identical experimental configurations at each exciton filling and extract the deviation between the two measurements. [Extended Data Fig. 9c](#) shows the results for three representative exciton fillings $v_{ex}$ = 0.02, 1.12 and 1.39, from which we obtain a standard deviation $\Delta\eta_{PL}$ of 0.21%, 0.17% and 0.17%. Such a small uncertainty corresponds to an error bar smaller than the symbol size in PL raw helicity

([Fig. 3b](#) and [Extended Data Fig. 9c](#)). On the other hand, the uncertainty in GH will be dramatically amplified at small $|\eta_{\text{pump}}|$ since GH = $\eta_{\text{PL}}/\eta_{\text{pump}}$, and GH becomes nominally ill-defined at $\eta_{\text{pump}} = 0$. We therefore extrapolate GH at $\eta_{\text{pump}} = 0$ from the limit of $\eta_{\text{pump}} \to 0$. As exemplified in [Extended Data Fig. 9d](#), the uncertainty in GH becomes reasonably small (<5%) when $|\eta_{\text{pump}}|>0.05$ (outside green shaded region), where the GH curve is already flat with pump polarization. This indicates that GH has a well-defined value in the $\eta_{\text{pump}} \to 0$ limit, and our extrapolation is reliable. Another way to understand the reliability of GH at small $|\eta_{\text{pump}}|$ is that it is simply the slope between $\eta_{\text{PL}}$ and $\eta_{\text{pump}}$. As one can directly see in the PL raw helicity ([Fig. 3b](#)), $\eta_{\text{PL}}$ has a well-defined slope near $|\eta_{\text{pump}}|=0$.

**Spin-½ Bose-Hubbard model:** To account for the two species of excitons related by time-reversal symmetry, the simplest form of the Bose-Hubbard model reads

$$H = \sum_{<i,j>,\alpha} -t b_{i,\alpha}^\dagger b_{j,\alpha} + h.c. + \sum_{i,\alpha} U(n_{i,\alpha} - 1/2)^2 + \sum_i V n_{i,1} n_{i,2}$$

Here the $\alpha=1,2$ label $K$ and $K'$ valley pseudo-spin of excitons, $t$ is hopping between nearest neighboring sites, and the interactions between excitons consist of intra-species repulsion $U$ and inter-species repulsion $V$. Our flavor-resolved pump-probe results indicate $U>V>0$, i.e., an on-site AFM interaction. Consequently, doublon sites always form between one $K$ valley and one $K'$ valley exciton, and the chemical potential jump at $v_{\text{ex}}=1$ directly measures $V\sim30$ meV.

On the other hand, the on-site interactions cannot account for the distinctive spin imbalance responses between LCP/RCP pump and linear pump at $v_{\text{ex}}>1$ ([Fig. 2](#)). The on-site AFM interaction can indeed induce different probe responses between different pump polarizations: an LCP/RCP pump will generate more $K/K'$ background excitons, which will suppress/assist the formation of doublon sites with the extra $K$ excitons from an LCP probe. Such effect should therefore be opposite under LCP and RCP pump compared to linear pump, which is incompatible with the symmetric behaviors of features in [Fig. 2](#) under both LCP and RCP pumps. In addition, the strength of the on-site interaction effects scales linearly with the doublon site density and should continuously increase with $v_{\text{ex}}$; while experimental features show sensitive and non-monotonic filling dependence. Last but not least, the effect from the AFM on-site interaction remain largely intact up to 60 K ([Extended Data Fig. 5](#)). However, all the observed features disappear at 60 K ([Fig. 5](#) and [Extended Data Fig. 6](#)). These pieces of evidence indicate a dominant role of inter-site spin interactions to features in Fig. 2-5.

**Theoretical phase diagram:** Inter-site spin interactions naturally emerge in the spin-½ Bose-Hubbard model. One well-established mechanism is the super-exchange effect $J_{SE}\sim t^2/V$ (see Supplementary Materials and Ref. [35]). In a Fermi Hubbard model, $J_{SE}$ is isotropically AFM. In a spin-½ Bose-Hubbard model, in contrast, its in-plane components $J_{SE,xy}$ are FM. Quantitatively, we estimate $J_{SE,xy}$ to be ~1 meV using typical hopping energy of moiré excitons

$t \sim 5$ meV (Ref. [6,57]) and the measured inter-species on-site interaction energy $V \sim 30$ meV. We can also independently estimate $J_{SE,xy} \sim 1$ meV from the temperature dependence using $6J_{SE,xy}=k_B T_c$, where $T_c=35$ K is the melting temperature of the $xy$ order. The consistency between two independent methods further confirms the validity of our estimation. The out-of-plane component of the super-exchange effect $J_{SE,z} \sim (2t^2/U - t^2/V)$ can be either FM or AFM[35]. Since $U>V>0$ guarantees $|J_{SE,z}|<|J_{SE,xy}|$, an FM-$xy$ order is expected at the Mott insulator state of $v_{ex}=1$. On the other hand, further doping of excitons will lead to additional effective FM interaction $J_N$ similar to Nagaoka ferromagnetism in Fermi Hubbard model[42,44] (see Supplementary Materials). While $J_N$ is isotropic in the Fermi Hubbard model with SU(2) spin rotation symmetry, the symmetry is explicitly broken in the spin-½ Bose-Hubbard model when $U \neq V$; and $J_N$ favors FM-$z$. As $J_N$ grows with exciton filling above $v_{ex}=1$, it eventually makes $J_z=J_{SE,z}+J_N>J_{SE,xy}$, leading to a transition between FM-$xy$ to FM-$z$ order. We encapsulate competition between the super-exchange effect and the Nagaoka-type ferromagnetism in a phenomenological XXZ model (see Supplementary Materials), which successfully captures all salient features of the experimental observation.

The two-orbital Bose-Hubbard model can potentially support a plethora of much more exotic phases beyond the ferromagnetic orders being discussed here. For example, it was shown numerically that a supersolid can be realized in the effective XXZ spin-1/2 model with one boson per-site[43]. In the Supplementary Materials we will also briefly discuss the potential exotic phases when we go beyond the vicinity of one boson per site, as two flavors of hard core bosons can be mapped to an SU(4) spin model (with anisotropies), which is a system that has attracted enormous interests in the past few decades as it can be engineered in transition metal oxides with both spin and orbital degrees of freedom, graphene-based moiré systems, as well as cold atoms[49–55]. Various exotic phases of spin systems with exact or approximate SU(4) symmetries have been discussed in literature[58–65].

**Discussions on magnetic field dependence:** The observed magnetic field dependence at $v_{ex}=1.1$ (Fig. 4a and b) can be intuitively understood through an interplay between the molecular field and the external field for an $xy$ pseudospin order. At zero external field, the $xy$ pseudospin order leads to finite $|\phi_x|$ and generates an in-plane mean field that aligns pseudospins to in-plane. With a sufficiently large external field, on the other hand, the total field becomes out-of-plane. As a result, $xy$ orders relying on the in-plane mean field are suppressed and $z$ orders are favored. Once the system enters the $z$ order ($|B_z|>20$ mT), the Zeeman field applied will not significantly break symmetry between the two pseudospins since the Zeeman energy scale (~0.1 meV at 20 mT) is much smaller than the exchange interaction (~1 meV). Indeed, we observe largely symmetric behaviors under positive and negative field for $|B_z|>20$ mT (Fig. 3e and Extended Data Fig. 7a). The intermediate field regime ($|B_z|<20$ mT) shows asymmetric responses for positive and negative $B_z$, indicating external symmetry breaking from the Zeeman field. As a result, there is no well-defined spontaneous symmetry breaking or pseudospin order. The asymmetric behaviors can be qualitatively understood from the fact that $|\phi_x|$ remains finite in this regime. The total field is therefore tilted and can still mix the two pseudospins and induce rapid switching between them, during which the Zeeman splitting favors the flavor with lower energy.

***xy* order**: The observed PL helicity is always zero at linear pump, indicating zero average out-of-plane "spin". There are two possible scenarios: all sites have in-plane "spin"; or domains of up and down spins. The second scenario applies to the case of $v_{ex} > 1.25$, which shows rapid increase of PL helicity with pump helicity and a "V" shape GH, as discussed in the main text. The distinctively different "Λ" shape GH at $v_{ex} \sim 1.1$ thus excludes this scenario and indicate in-plane "spin". In addition, the symmetric and sensitive magnetic field dependence ([Fig. 2d-f](#)) indicates a finite in-plane mean field at linear pump. Therefore, the in-plane spin directions cannot be completely random and the system should have at least a finite-range, transient *xy* order. On the other hand, our observation does not require a long-range *xy* order and provides no information on the correlation length of the order. A long-range FM-*xy* order corresponds to a global phase coherence between the two valleys; therefore, PL from the system should be linearly polarized. We did not observe linear helicity in PL ([Extended Data Fig. 8](#)), indicating that there is no true long-range FM-*xy* order. This can be naturally understood as there can only be at most a quasi-long range *xy* order in 2D at finite temperature[66]; and all orders observed in this work should be of transient nature. Besides intrinsic spin fluctuations and exciton decay, coupling to phonons and other quasiparticles may further introduce decoherence channels acting as a random in-plane magnetic field, thereby limiting the time- and length-scale of the *xy* order.

**Acknowledgement:** C.J. acknowledges support from Air Force Office of Scientific Research under award FA9550-23-1-0117. R.X. acknowledges support from the UC Santa Barbara NSF Quantum Foundry funded via the Q-AMASE-i program under award DMR-1906325. K.W. and T.T. acknowledge support from the JSPS KAKENHI (Grant Numbers 19H05790 and 20H00354). S.T acknowledges primary support from DOE-SC0020653 (materials synthesis), Applied Materials Inc., NSF CMMI 1825594 (NMR and TEM studies), NSF DMR-1955889 (magnetic measurements), NSF CMMI-1933214, NSF 1904716, NSF 1935994, NSF ECCS 2052527, DMR 2111812, and CMMI 2129412. C. X. is supported by the Simons Investigator program.

**Author contribution:** C.J. conceived and supervised the project. R.X., J.H.N. and Z.Z. fabricated the devices. R.X. performed the optical measurements. R.X. and S.L.B analyzed the data. K.S. and C.X. performed theoretical calculations on the spin model. R.B., H.R. and S.T. grew the WSe$_2$ and WS$_2$ crystals. K.W. and T.T. grew the hBN crystals. C.J. and R.X. wrote the manuscript with the input from all the authors.

**Competing interests:** The authors declare no competing interests.

**Data availability:** All data supporting this work are available from the corresponding author upon request.

Figures:

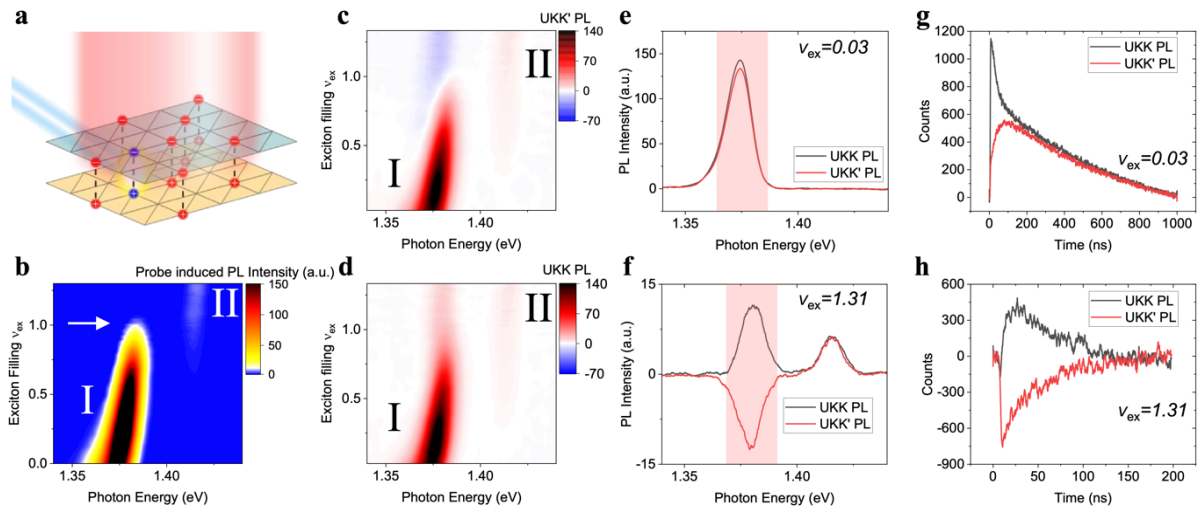

**Fig. 1: Spin-1/2 Bose-Hubbard model a,** Schematics of our pump-probe spectroscopy on a type-II hetero-bilayer $WSe_2/WS_2$. The background interlayer-exciton density (red) is controlled by the pump light and the charge density is kept at zero. The probe light injects an extra interlayer-exciton (blue), whose response is isolated through lock-in detection. **b,** Exciton-filling dependence of probe-induced PL spectrum using unpolarized pump and probe light. A sudden jump of exciton chemical potential at $v_{ex}=1$ is observed (white arrow), indicating an incompressible state of exciton. Low (high) energy emission peak is denoted as peak I (II). **c,d,** Polarization-resolved probe-induced PL spectra as a function of exciton filling. A linear pump light is used to generate equal numbers of $K$ and $K'$ valley excitons in the background, while an LCP probe light selectively excites extra $K$ valley excitons. $K'$ valley (**c**) or $K$ valley (**d**) PL response induced by the probe light is collected separately. UKK (UKK') refers to pump injecting unpolarized excitons, probe injecting $K$ valley excitons and PL detecting $K$ ($K'$) valley excitons. **e,f,** Linecuts of (**c**) and (**d**) at low (**e**) and high (**f**) exciton density. **g,h,** Probe-induced TRPL signals from peak I at low (**g**) and high (**h**) exciton density. A constant background signal from the CW linear pump light is subtracted. Pink boxes in (**e**) and (**f**) denote the spectral filter used to isolate peak I response. The negative signals in UKK' configuration (**h**) indicate that $K$ excitons selectively form doublons with $K'$ excitons.

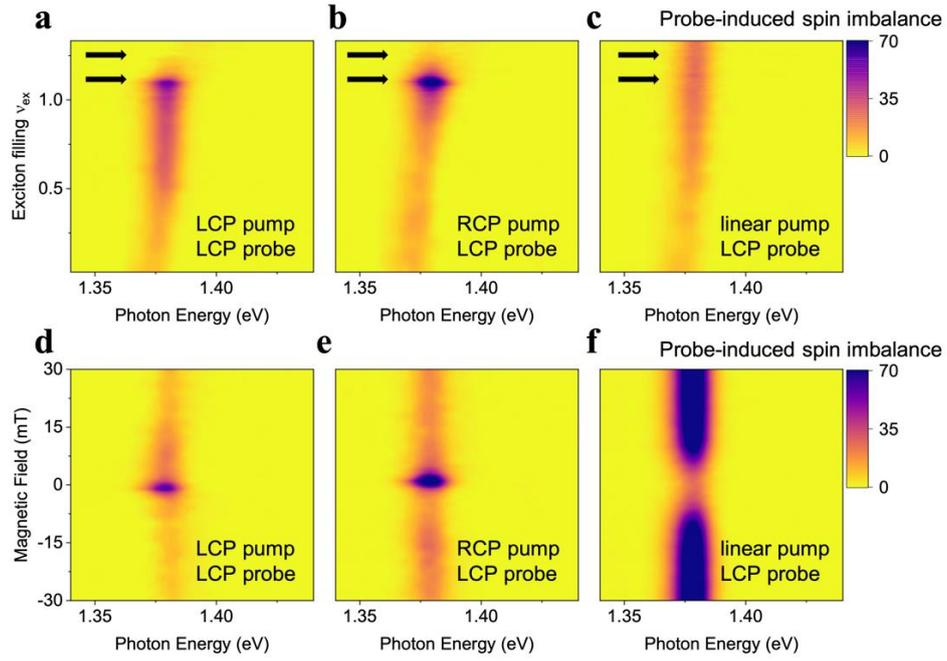

**Fig. 2: Inter-site spin-dependent exciton interaction. a-c,** Probe-induced spin imbalance as a function of background exciton fillings for LCP (**a**), RCP (**b**) and linear (**c**) pump respectively. For both LCP and RCP pump, a sharp enhancement of signals at $\nu_{ex}\sim1.1$ followed by a quick drop at $\nu_{ex}\sim1.2$ are observed (black arrows). In contrast, these features are missing under linear pump, indicating rapidly changing exciton spin interaction when doping slight away from the correlated insulator state. **d-f,** Evolution of probe-induced spin imbalance at $\nu_{ex}\sim1.1$ under out-of-plane magnetic field $B_z$ for LCP (**d**), RCP (**e**) and linear (**f**) pump, respectively. The signals show sensitive and symmetric change under a small $|B_z|\sim5$ mT. All the measurements are performed at base temperature 3 K unless specified.

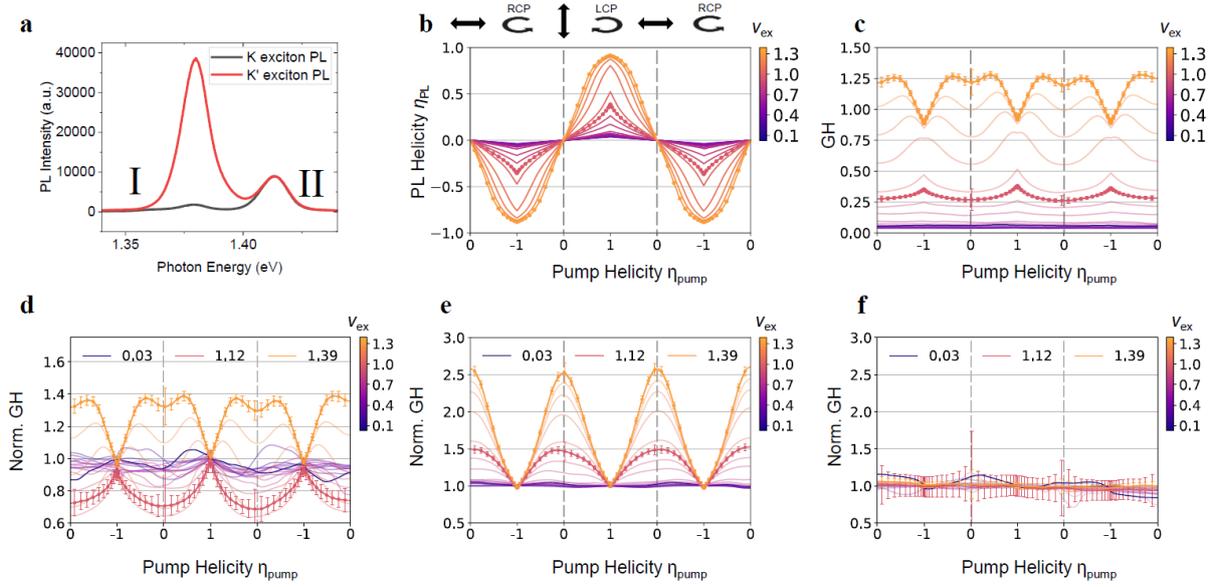

**Fig. 3: Evidence of exciton spin orders from generalized helicity (GH). a,** pump-only PL from K (black) and K' (red) valley with RCP pump at $v_{ex}$ = 1.39. Peak I shows a large PL helicity while peak II has no PL helicity. **b-d,** PL raw helicity $\eta_{PL}$ (**b**), generalized helicity (GH, defined as $\eta_{PL}/\eta_{pump}$) (**c**) and normalized GH (**d**) of peak I under different pump helicity $\eta_{pump}$ and exciton fillings $v_{ex}$. GH does not depend on $\eta_{pump}$ at low exciton density, consistent with the single-particle picture with no spin order. In contrast, GH becomes "Λ" shape at $v_{ex}$~1.1 and quickly transitions into "V" shape at $v_{ex}$>1.25, indicating existence of mean-field from exciton spin order. **e,** Normalized GH at $B_z$ = -30 mT (see $B_z$ = 30 mT in Extended Data Fig. 7a). The "Λ" shape at $v_{ex}$~1.1 becomes "V" shape under ±30 mT field. The sensitive and symmetric $B_z$ dependence echoes with the pump probe measurement results. **f,** Normalized GH at 60 K. GH remains flat over the whole exciton filling range. GH and normalized GH at $v_{ex}$ = 0.03, 1.12, 1.39 are highlighted in **c-f.** Data are shown as lines and symbols for $v_{ex}$ = 1.12 and 1.39; and only lines are shown at other fillings for visual clarity. Error bars represent standard deviation in PL helicity, GH and normalized GH (**b-f**, see Methods: Data analysis).

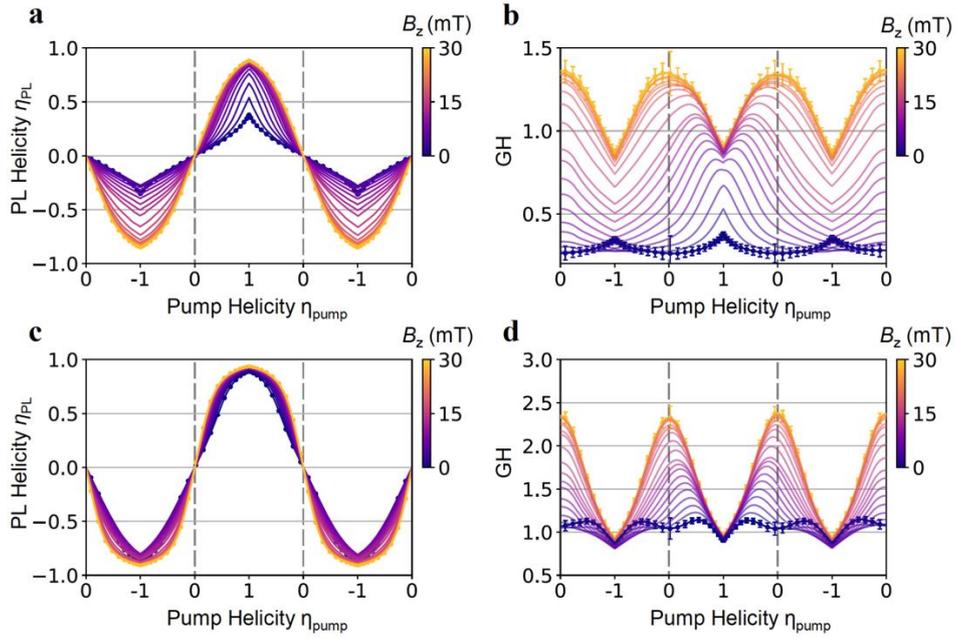

**Fig. 4: Tunable exciton spin orders with magnetic field and exciton filling. a,b,** Magnetic field dependence of PL raw helicity $\eta_{PL}$ (**a**) and GH (**b**) at $v_{ex}$=1.1 from 0 to 30 mT. The GH transforms from a "Λ" shape to a "V" shape and saturates at ~20 mT, suggesting a phase transition. At high field, GH exceeds 1 in a wide range of pump helicity, indicating spontaneous increase of spin polarization, which is the hallmark of an FM-$z$ spin order. **c,d,** Same as **a,b** for $v_{ex}$=1.3. The zero field GH is similar to the high field GH at $v_{ex}$=1.1, signifying an FM-$z$ spin order at zero field. The "V" shape GH is further enhanced by $B_z$. Data are shown as lines and symbols for $B_z$ = 0 and 30 mT; and only lines are shown at other magnetic fields for visual clarity. Error bars in **a** and **c** are smaller than the symbol size.

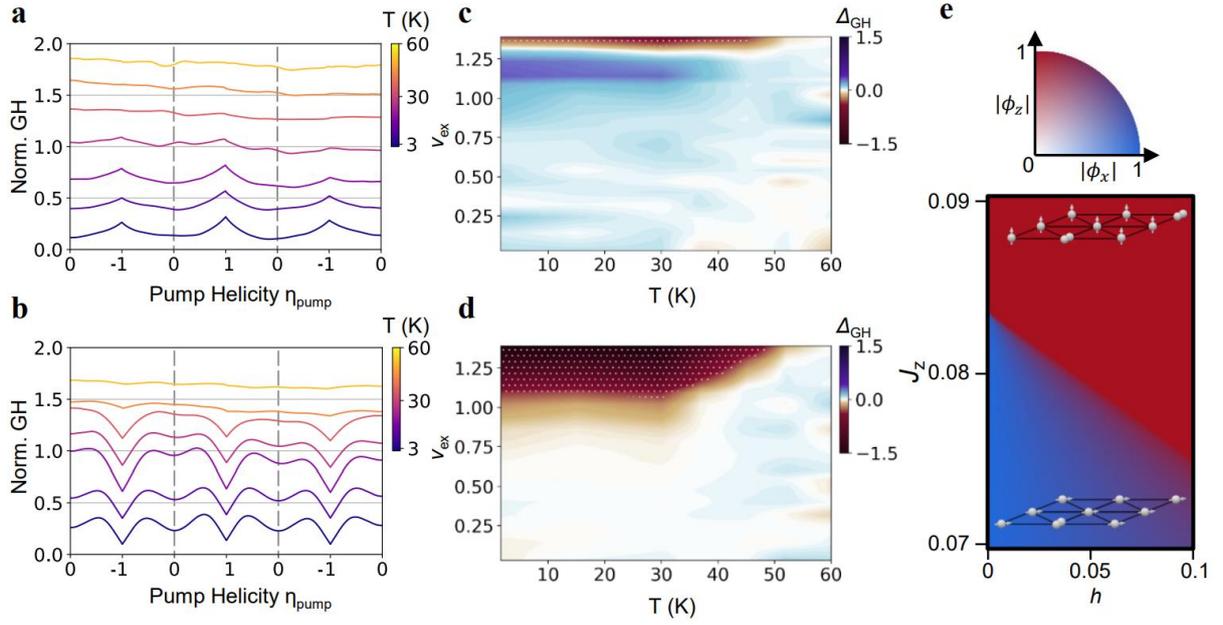

**Fig. 5: Phase diagram of exciton spin orders. a,b,** Temperature dependence of normalized GH at $v_{ex}$=1.1 (**a**) and $v_{ex}$=1.3 (**b**). The "Λ" shape and "V" shape feature melt at around 35 K and 50 K, respectively. **c,d,** Phase diagrams of $\Delta_{GH}$ at $B_z$=0 mT (**c**) and -30 mT (**d**). A positive (negative) $\Delta_{GH}$ corresponds to a "Λ" ("V") shape GH and indicates $xy$ ($z$) spin order. White dotted texture marks regions with GH >1 at linear pump, which is the hallmark of an FM-$z$ spin order. **e,** Theoretical phase diagram from a phenomenological spin ½ XXZ model, where $h$ is the out-of-plane magnetic field and $J_z$ is $z$-direction exchange interaction. The in-plane exchange $J_\perp$ is fixed to be 1/6. $\phi_x$ ($\phi_z$) is the expectation value of $S^x$ ($S^z$). The color represents orientation of the order parameter and the opacity represents its amplitude. Effects from adding extra excitons to a $v_{ex}$=1 correlated insulator is captured by $J_z$ that increases with exciton filling. A transition from the FM-$xy$ to FM-$z$ order is expected upon both increasing $J_z$ (exciton filling) and magnetic field, which is consistent with our experimental observations.

**Extended Data Figures:**

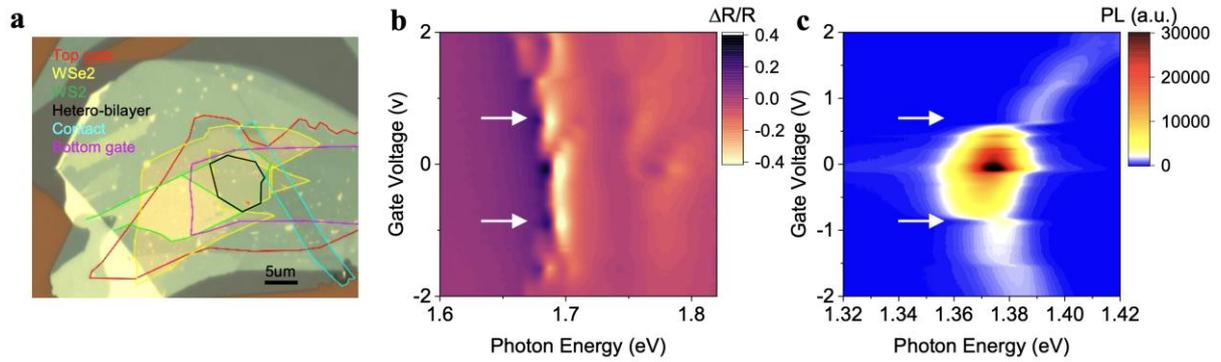

**Extended Data Fig. 1: Basic characterizations of devices. a**, optical image of a representative dual-gated 0-degree-aligned WSe$_2$/WS$_2$ device D1. Yellow and green solid lines denote contours of the monolayer WSe$_2$ and WS$_2$ flakes, respectively. **b,c,** Electron doping-dependent absorption (**b**) and PL (**c**) spectrum of device D1 at zero pump intensity. White arrows indicate one filling of electrons and holes.

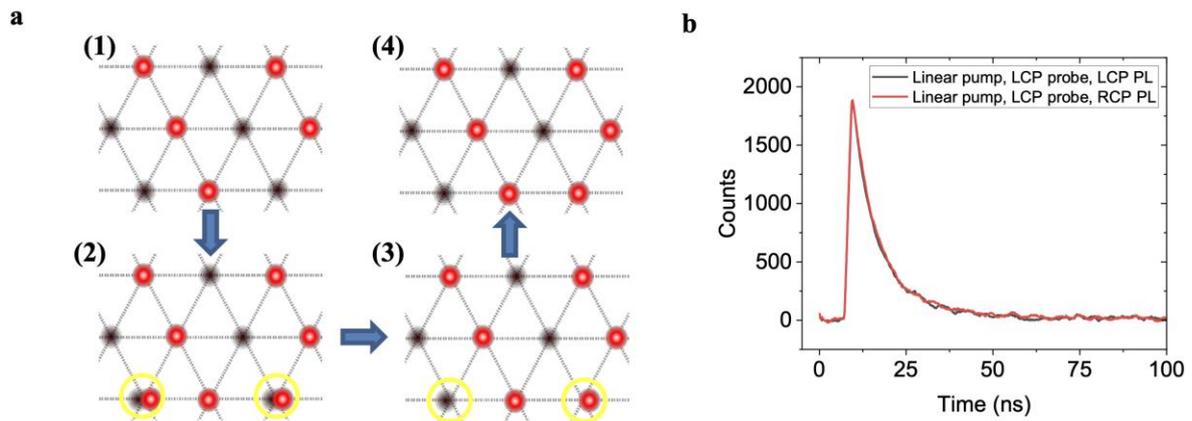

**Extended Data Fig. 2: Response of a bosonic correlated insulator to transient extra *K* excitons. a, (1),** CW linear pump light injects equal number of *K* valley (red) and *K'* valley (grey) background excitons that form an exciton lattice. **(2),** Pulsed circular polarized probe light transiently injects two extra *K* valley excitons, which takes two *K'* sites to form doublon sites (yellow circles). **(3),** Doublon sites have equal probabilities to emit *K(K')* excitons and leave a single site of *K'(K)*. **(4),** After the doublons decay the system will have one more *K* single site and one less *K'* single site, giving rise to negative *K'* response and positive *K* response of peak I with equal amplitude. **b,** Probe-induced TRPL measurements of doublon emissions under linear pump and LCP probe configuration. The two valleys show identical amplitude and dynamics, consistent with the expectation of doublon emission.

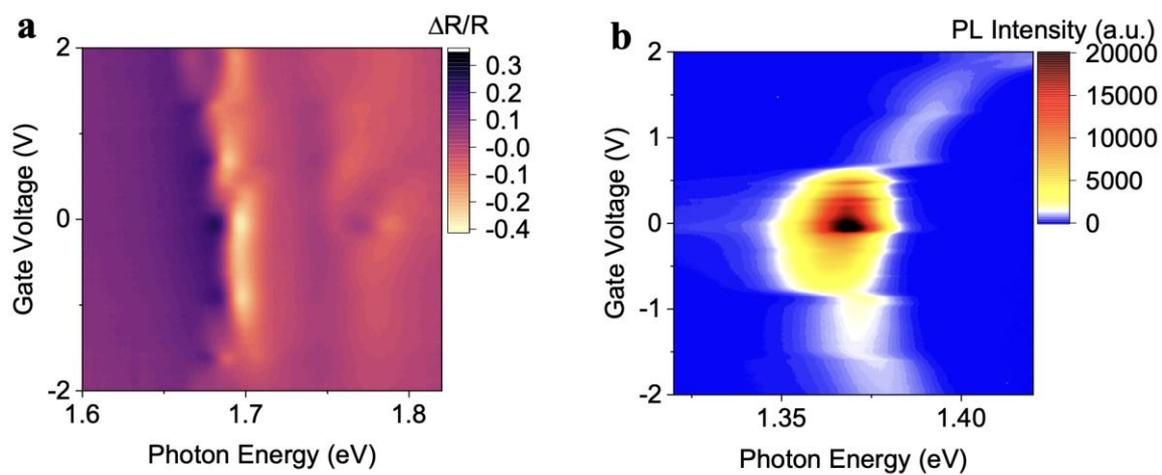

**Extended Data Fig. 3: Optical characterization of device D2.** Doping-dependent absorption (**a**) and PL (**b**) spectrum of the 0-degree aligned $WSe_2/WS_2$ device D2 at zero pump intensity.

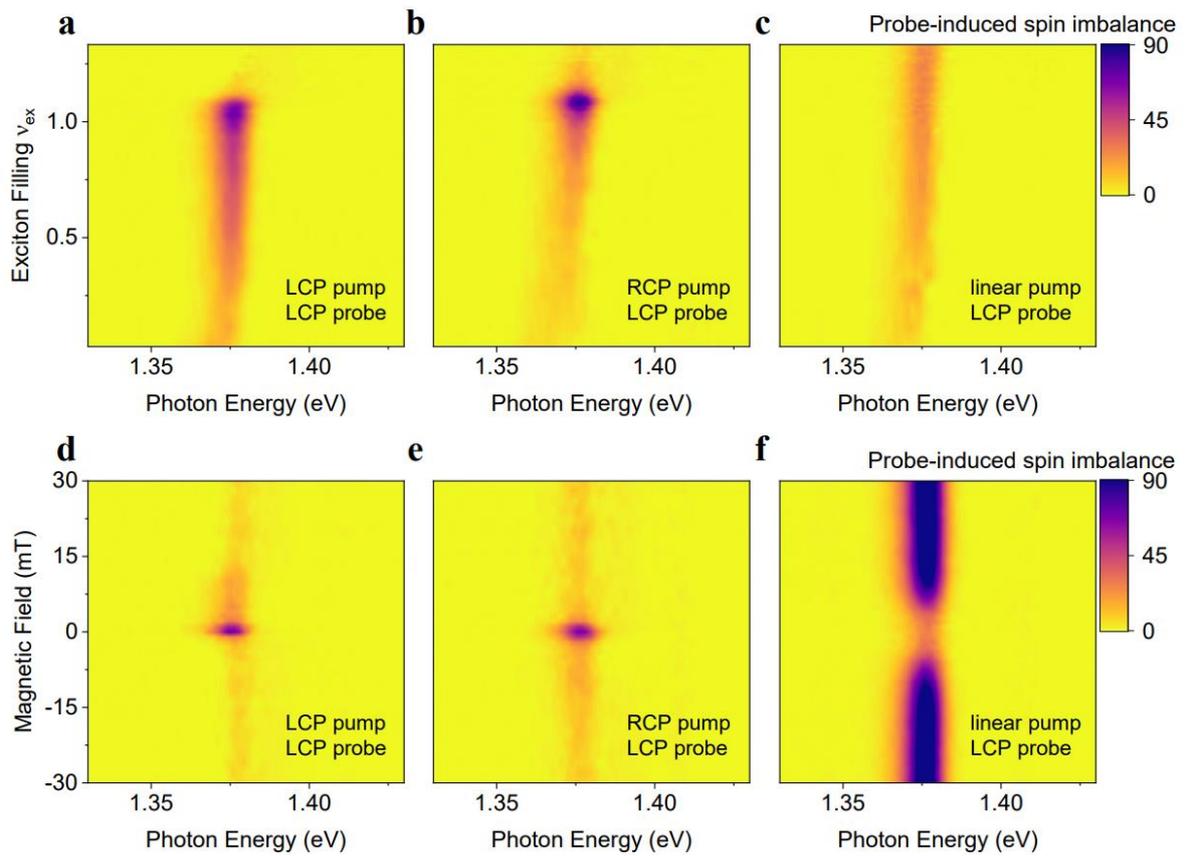

**Extended Data Fig. 4: Probe-induced spin imbalance in device D2. a-c,** LCP probe-induced spin imbalance as a function of background exciton fillings for LCP (**a**), RCP (**b**) and linear (**c**) pump respectively. **d-f,** Evolution of spin imbalance signal at $\nu_{ex}$=1.1 under out-of-plane magnetic field $B_z$ for LCP (**d**), RCP (**e**) and linear (**f**) pump, respectively.

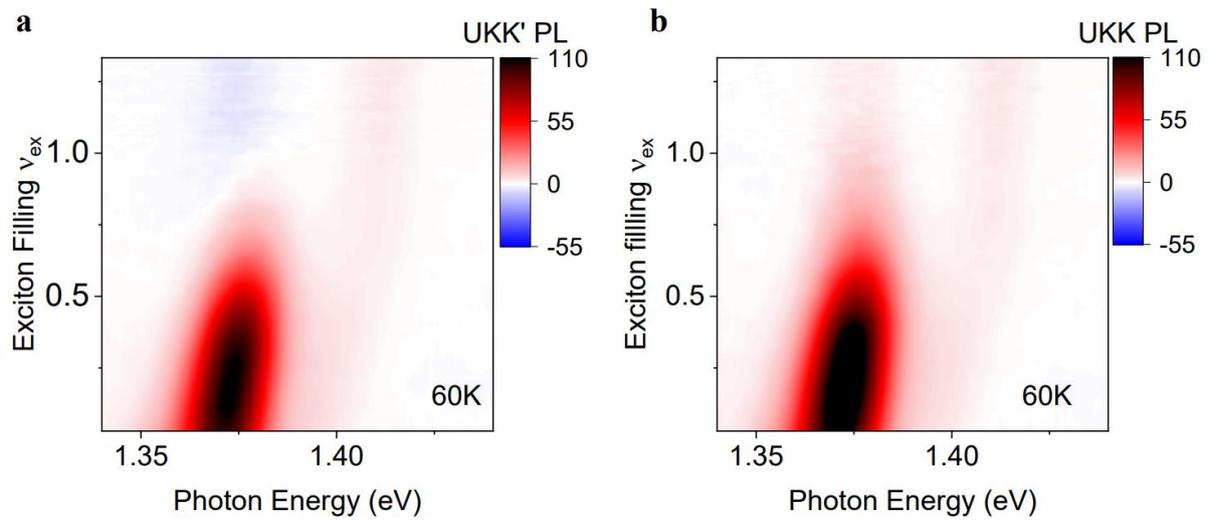

**Extended Data Fig. 5: Polarization-resolved probe-induced PL spectra of device D1 at 60 K.** The bosonic correlated insulator state and on-site AFM interaction are still robust at 60 K. UKK (UKK') refers to pump injecting unpolarized excitons, probe injecting $K$ valley excitons and PL detecting $K$ ($K'$) valley excitons.

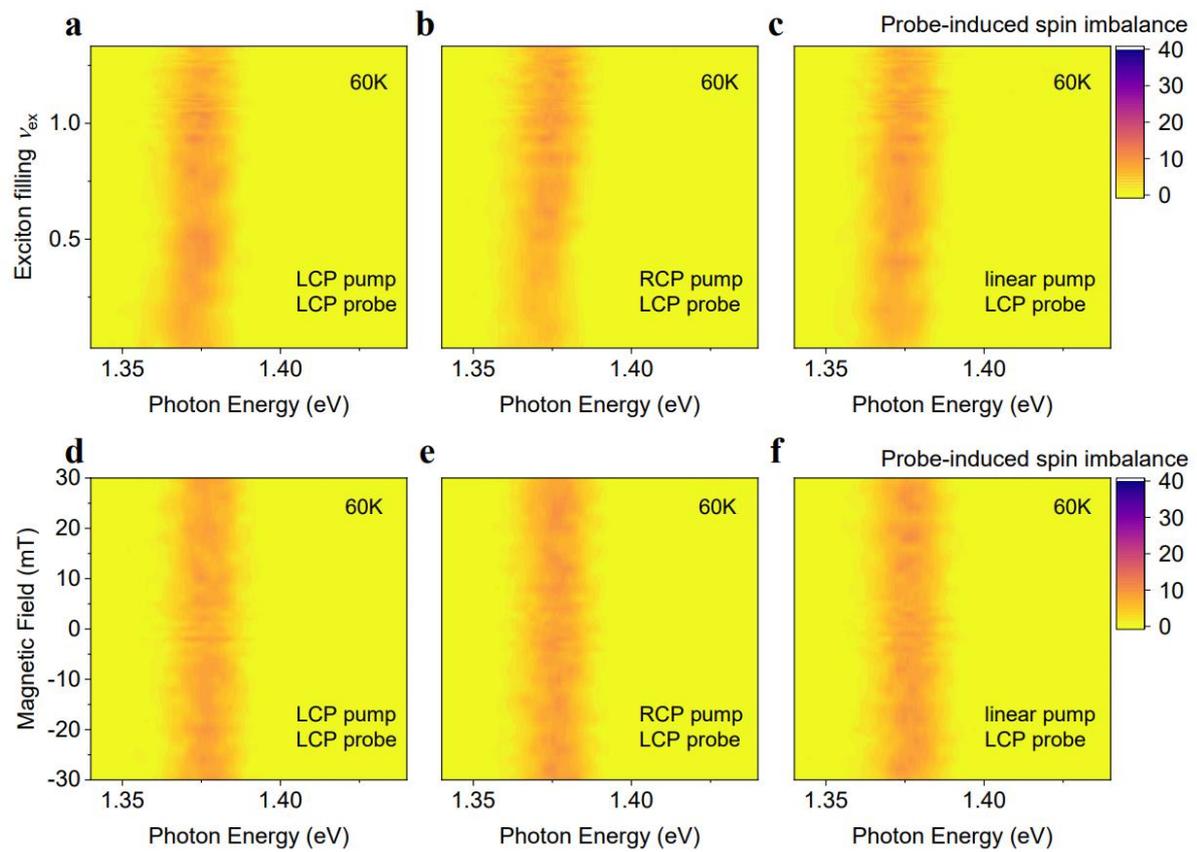

**Extended Data Fig. 6: Probe-induced spin imbalance in device D1 at 60 K.** No pump polarization dependence is observed across all exciton fillings, indicating vanishing spin-dependent interactions at 60 K.

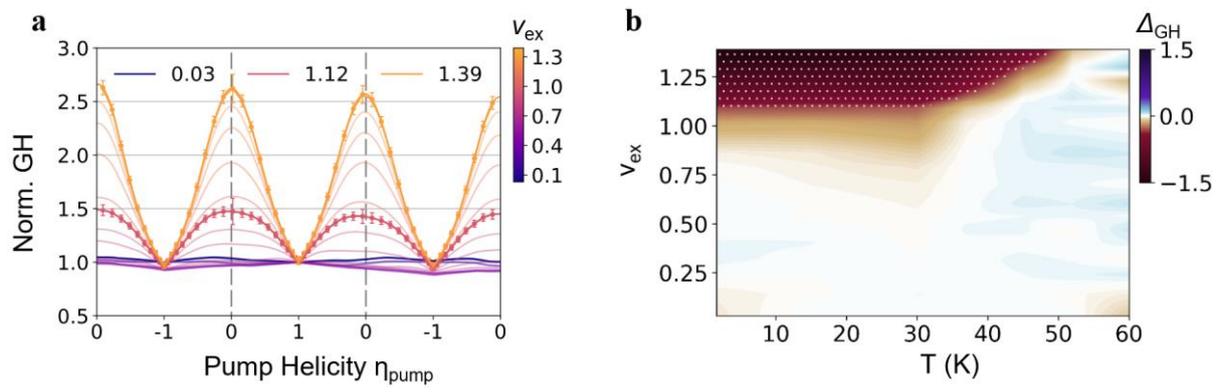

**Extended Data Fig. 7: Normalized generalized helicity and phase diagrams of $\Delta_{GH}$ at $B_z$= 30 mT**.

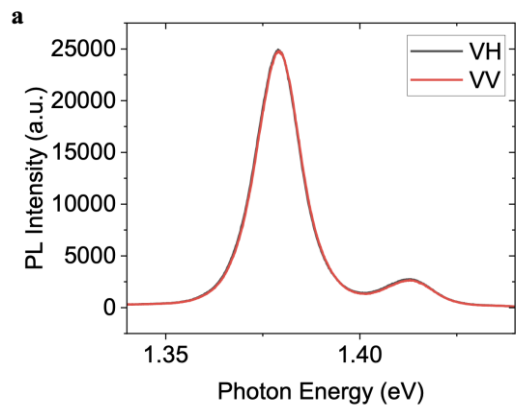

**Extended Data Fig. 8: Linear polarization-resolved PL at $v_{ex} = 1.1$.** No linear helicity is observed between vertical (VV) and horizontal (VH) PL detection, indicating that there is no global long-range FM-$xy$ order. The pump light is linearly polarized along the vertical direction.

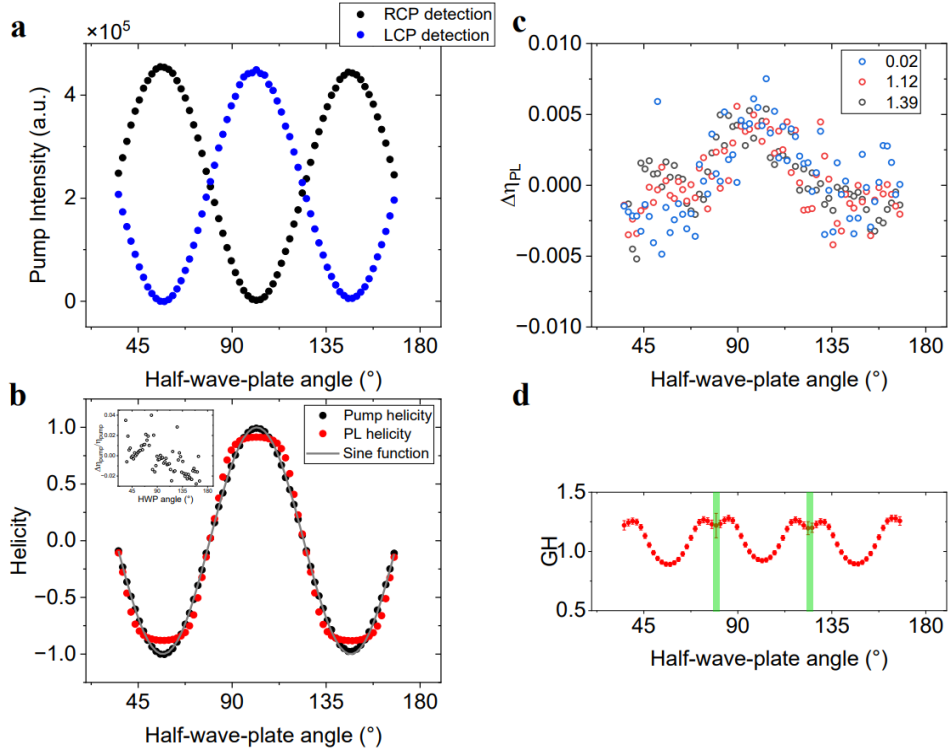

**Extended Data Fig. 9: Estimation of measurement uncertainties. a,** LCP (blue) and RCP (black) components of the sample-reflected pump light under identical experimental configuration as polarization-resolved PL measurements. **b,** Pump helicity (black symbols) and PL helicity (red symbols) at $v_{ex}$=1.39 and zero magnetic field with different HWP angles. The pump helicity shows near-perfect match with theoretical curve (grey line) with a relative standard deviation of 1.7%. Inset shows the deviation between the measured and theoretical pump helicity. **c,** The deviation in $\eta_{PL}$ between two successive measurements, from which we calculate the standard deviation in $\eta_{PL}$ to be 0.21%, 0.17% and 0.17% for $v_{ex}$=0.02, 1.12 and 1.39, respectively. **d,** Generalized helicity (GH) at $v_{ex}$=1.39 and zero magnetic field as a function of HWP angles. The standard deviation in GH becomes reasonably small (<5%) when $|\eta_{pump}|$>0.05 (outside of the green shaded region).

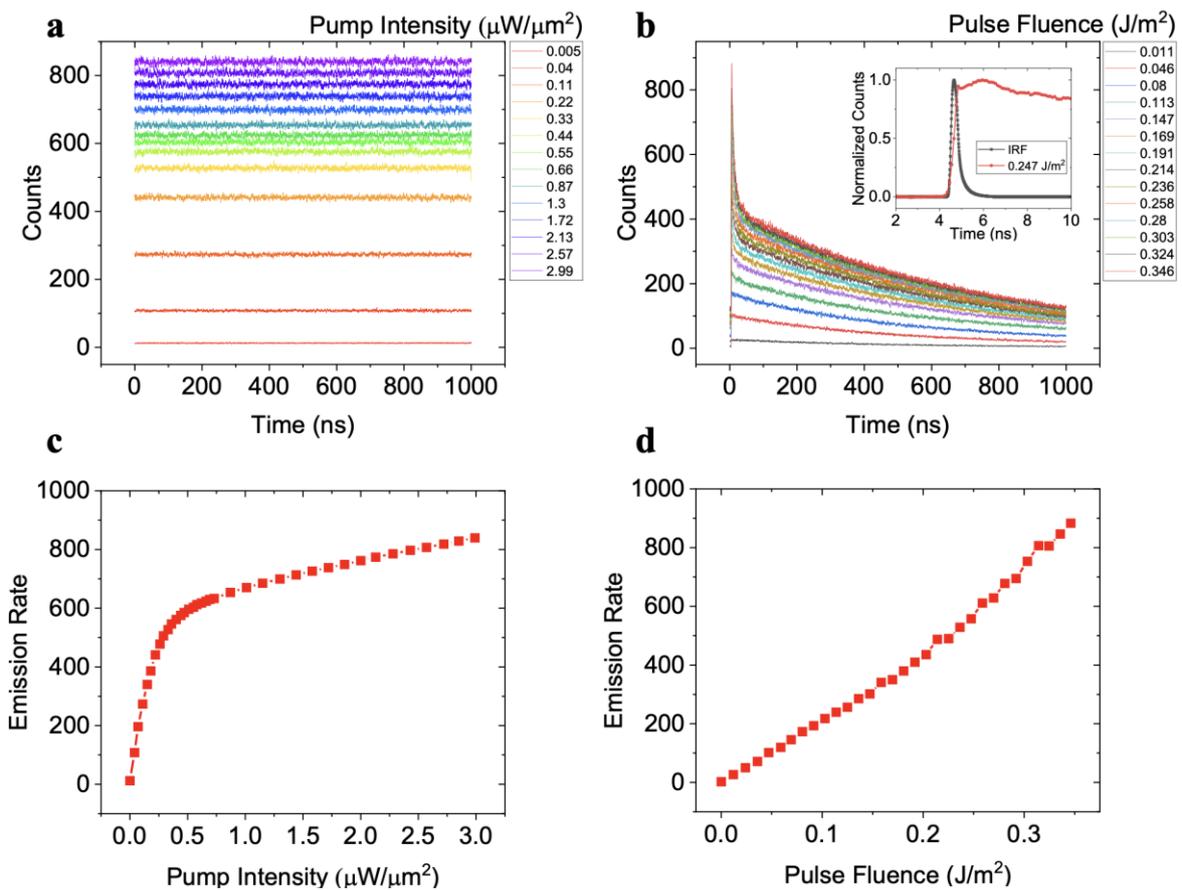

**Extended Data Fig. 10: Calibration of exciton density via TRPL measurement.** **a,** TRPL using a 660 nm CW pump light of different pump intensity. **b,** Same as **a** but with a 660 nm pulsed pump light (300 ps duration, 1 MHz repetition rate). Inset: Comparison between IRF and PL dynamics indicates negligible exciton relaxation immediately after time zero. **c,** Emission rates from the CW pump light of different intensities. **d,** Emission rates from the pulsed pump light of different fluences.

# Supplementary Material : Mean Field analysis of a phenomenological spin model

## I. HAMILTONIAN ON THE MOIRE SUPERLATTICE

We would like to first write down the Hamiltonian for interacting excitons on the moire superlattice. The symmetry of the moire lattice should include $P_x$ which takes $x \to -x$, time-reversal $\mathcal{T}$, and three fold rotation $R_{2\pi/3}$. The $P_x$ and $\mathcal{T}$ both interchanges the two species (pseudo-spin flavors) of the excitons, as the exciton is formed with electron-hole pair from the $K$ and $K'$ valleys respectively. The symmetry allowed Hamiltonian reads

$$H = \sum_{<i,j>,\alpha} -t e^{i\phi_{ij}\tau^\alpha} b^\dagger_{i,\alpha} b_{j,\alpha} + H.c. + \sum_{i,\alpha} U(\hat{n}_{i,\alpha} - 1/2)^2 + \sum_i V \hat{n}_{i,1} \hat{n}_{i,2} \cdots \quad (1)$$

The interactions between the excitons include an usual intra-species repulsion $U$ as well as an inter-species repulsion $V$. Naturally we expect $U > V > 0$.

The symmetry of the system allows the two speciecies of excitons to see opposite fluxes through each triangular plaquette of the moire lattice, and the phase angle of the hopping amplitudes of the excitons satisfy $\tau^1 = -\tau^2 = \pm 1$, and $\phi_{ij}$ changes sign under rotation $\pi/3$, hence $\phi_{ij}$ would vanish if the system had a six fold rotation symmetry. The fluxes of the excitons can affect the effective pseudo-spin model arising from superexchanges. If there were no flux, it is well-known that, at precisely filling $\nu_{ex} = 1$ (one exciton boson per moire unit cell), the pseudo-spin physics of the system is captured by the following XXZ model in the limit $U > V \gg t$ [1]:

$$H = -\sum_{<ij>} J_\perp (S^x_i S^x_j + S^y_i S^y_j) - \sum_{<ij>} J_z S^z_i S^z_j \quad (2)$$

here $J_\perp$ is positive, $J_z$ is negative, i.e. the system has a ferromagnetic inplane spin interaction, and an antiferromagnetic interaction between $S^z$. The pseudo-spin operators correspond to the boson operators in the following way:

$$S^+_i \sim b^\dagger_{i,2} b_{i,1}, \quad S^-_i \sim b^\dagger_{i,1} b_{i,2}, \quad S^z_i \sim \hat{n}_{i,2} - \hat{n}_{i,1}. \quad (3)$$

The flux in the pseudo-spin Hamiltonian will turn on certain amount of frustration for the inplane components of the pseudo-spin, which is analogous to the fluxes seen by the orbital degrees of freedom derived for some of the graphene-based moire systems [2, 3].

## II. A PHENOMENOLOGICAL MODEL

The exact phase diagram of the Bose-Hubbard model (with fluxes), especially its fate at different fillings, deserve serious numerical studies. In this section we will discuss the physics of the system under doping away from $\nu_{ex} = 1$ with a simple phenomenological pseudo-spin model. Our phenomenological model Hamiltonian is motivated by the following observed phenomena of the exciton physics in TMD moire heterostructure:

(1) Near exciton filling $\nu_{ex} = 1$, the pseudo-spin-1/2 degree of freedom of the exciton is in an inplane ferromagnet phase (labelled as the FM-$xy$ phase); when doped with extra excitons, the pesudo-spin of the excitons is in the out-of-plane ferromagnet phase (labelled as the FM-$z$ phase).

(2) Near filling $\nu_{ex} \sim 1.1$, the pseudo-spin is sensitive to a weak external Zeeman field: the pseudo-spin polarization $\langle S^z \rangle$ rapidly saturates under the external Zeeman field.

These phenomena can be qualitatively understood as a competition between the superexchange effect, and doping with extra excitons. First of all, as we have discussed, for a Mott insulator of bosons with pseudo-spin-1/2 internal degree of freedom, the superexchange which arises from virtual hopping of bosons will yield a *ferromagnetic* interaction between inplane components of the pseudo-spins (here we take $\phi_{ij} = 0$), and an *antiferromagnetic* interaction between $S^z$. It was also shown numerically that for a broad range of parameters for the XXZ spin-1/2 model on a triangular lattice, the system would have a FM-$xy$ order [4] when the inplane superexchange is ferromagnetic. This is consistent with the observed inplane FM-$xy$ phase at $\nu_{ex} = 1$.

In the following we argue that, under doping of extra bosons, the kinetic energy of the extra boson density would favor a FM-$z$ order. This effect is based on the natural assumption that the intra-species onsite repulsion between the bosons is far stronger than the inter-species repulsion, and we also assume that the on-site repulsion interactions are far greater than the hopping amplitude of the excitons. The system would form a strong Mott insulator at filling $\nu_{ex} = 1$, and in the strong repulsion limit all the pseudospin configurations are degenerate. But this degeneracy is lifted under doping with one extra boson, as for a doped boson to hop freely, it is favorable for all the "background" bosons at filling $\nu_{ex} = 1$ to have the same polarization of pseudo-spin $S^z$, say $S^z = +1$, while the extra boson has the opposite pseudo-spin $S^z = -1$, since as we assumed, the intra-species repulsion is far stronger than the inter-species repulsion. Hence the kinetic energy of the doped bosons would favor the system to have a net ferromagnetic polarization along the $z$ direction. The physics here is to some extent analogous to the well-known doping-induced ferromagnet, i.e. the so-called Nagaoka ferromagnetism. We note here that the original Nagaoka's ferromagnet for the fermionic Hubbard model is isotropic in the spin space, but there is no SU(2) spin rotation symmetry for the pseudo-spin degree of freedom of the excitons. The SU(2) symmetry is broken explicitly by the difference between the intra-species and the inter-species repulsion, and the ferromagnetism for bosons under doping would favor a FM-$z$ order.

In order to qualitatively capture the interpolation between the FM-$xy$ and FM-$z$ phases, we design the following phenomenological spin-1/2 Hamiltonian on the triangular moire superlattice:

$$H = -\sum_{<ij>} J_\perp (S_i^x S_j^x + S_i^y S_j^y) - \sum_{<ij>, \ll ij\gg} J_z S_i^z S_j^z - \sum_i h S_i^z. \tag{4}$$

The nearest-neighbor inplane interaction $J_\perp > 0$ (ferromagnetic) captures the effect of the superexchange at $\nu_{ex} = 1$. Since our experiments mostly probes the pseudo-spin physics of the exciton, we encapsulates the effects of the doped excitons into the renormalization of $J_z$. $J_z$ would be antiferromagnetic ($J_z < 0$) at filling $\nu_{ex} = 1$, but it will gradually evolve into ferromagnetic interaction ($J_z > 0$) under doping. In our model Hamiltonian we include both first and second neighbor interactions for $J_z$, this is because if there were only nearest neighbor interaction between $S^z$, there would be a SU(2) symmetry at $J_\perp = J_z$, which as we commented before is unphysical. The mean field analysis to be presented here should be qualitatively independent of the microscopic details of the model.

We follow the standard procedure of mean field theory, and decompose the Hamiltonian as follows:

$$\begin{aligned} H_{MF} = &- \sum_{<ij>} J_\perp (\phi_x S_j^x + S_i^x \phi_x - \phi_x^2 + \phi_y S_j^y + S_i^y \phi_y - \phi_y^2) \\ &- \sum_{<ij>, \ll ij\gg} J_z (\phi_z S_j^z + S_i^z \phi_z - \phi_z^2) - \sum_i h S_i^z \end{aligned} \tag{5}$$

$\phi_a$ can be physically viewed as the expectation value of $S^a$. The partition function of the mean field theory at finite temperature is given by

$$\begin{aligned} Z &= \text{Tr}\{\exp{(-\beta H_{MF})}\} \\ &= \prod_i \text{Tr}\{\exp\{\beta(6J_\perp \phi_x S^x + 6J_\perp \phi_y S^y + 12 J_z \phi_z S^z + h S^z\}\} \\ &\times \exp\beta(-\sum_{<ij>} J_\perp \phi_x^2 + J_\perp \phi_y^2 - \sum_{<ij>, \ll ij\gg} J_z \phi_z^2) \\ &= \prod_i \cosh\beta\sqrt{(6J_\perp \phi_x)^2 + (6J_\perp \phi_y)^2 + (12 J_z \phi_z + h)^2} \times \exp\left(-\beta(3N J_\perp (\phi_x)^2 + 3N J_\perp (\phi_y)^2 + 6N J_z (\phi_z)^2)\right) \\ &= \left(\cosh\beta\sqrt{(6J_\perp \phi_x)^2 + (6J_\perp \phi_y)y^2 + (12 J_z \phi_z + h)^2} \exp\left(-\beta(3 J_\perp \phi_x^2 + 3 J_\perp \phi_y^2 + 6 J_z \phi_z^2)\right)\right)^N \end{aligned} \tag{6}$$

The O(2) symmetry in the XY plane allows us to just consider $\phi_x$ and set $\phi_y = 0$ without loss of generality. The expression of the partition function then leads to the following free energy:

$$\begin{aligned} \frac{F}{N} &= -\frac{1}{\beta N} \ln Z \\ &= -\frac{1}{\beta} \ln\left(\cosh\beta\sqrt{(6J_\perp \phi_x)^2 + (12 J_z \phi_z + h)^2}\right) + 3 J_\perp \phi_x^2 + 6 J_z \phi_z^2 \end{aligned} \tag{7}$$

The variational condition would give us the following consistency equations to determine $\phi_a$:

$$\frac{\partial F}{\partial \phi_x} = 0 \quad \rightarrow \quad \frac{6 J_\perp \phi_x \tanh\left(\beta\sqrt{(6J_\perp \phi_x)^2 + (12 J_z \phi_z + h)^2}\right)}{\sqrt{(6J_\perp \phi_x)^2 + (12 J_z \phi_z + h)^2}} = \phi_x \tag{8}$$

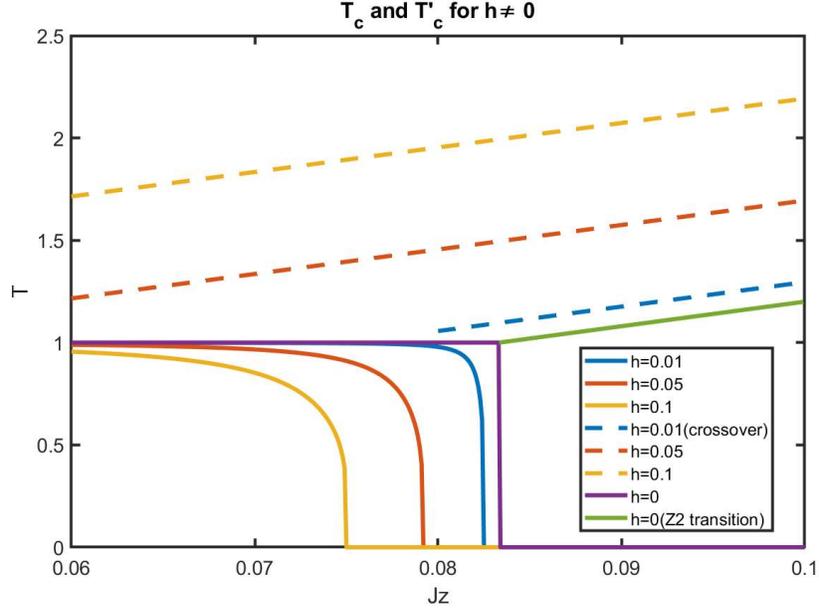

FIG. 1. A plot of $T_c$ and $T'_c$ for different $h$.

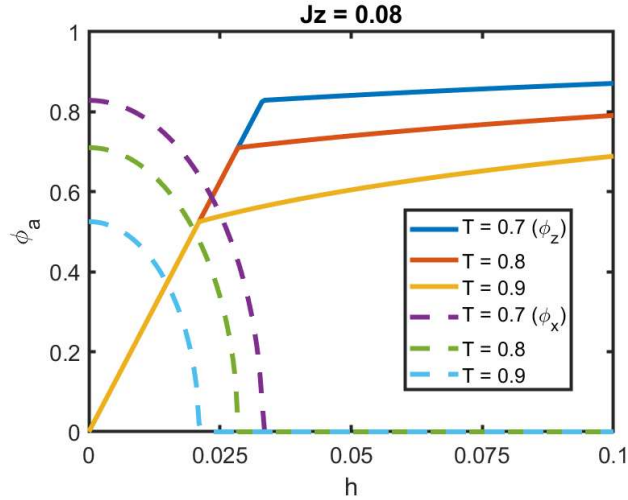

FIG. 2. A plot of $\phi_z$ and $\phi_x$ vs h

$$\frac{\partial F}{\partial \phi_z} = 0 \quad \rightarrow \quad \frac{(12J_z\phi_z + h)\tanh\left(\beta\sqrt{(6J_\perp\phi_x)^2 + (12J_z\phi_z + h)^2}\right)}{\sqrt{(6J_\perp\phi_x)^2 + (12J_z\phi_z + h)^2}} = \phi_z \qquad (9)$$

The phase diagram of the mean field theory is plotted in Fig. 1. For small $J_z$, the system at zero temperature has a long range FM-$xy$ order; at finite temperature the FM-$xy$ order will become a quasi-long range order with power-law correlation. Within the mean field theory, the critical temperature of the FM-$xy$ order is given by $T_c = 6J_\perp$, hence we choose $J_\perp = 1/6$ in the phase diagram to fix the energy scale of the superexchange interaction. While increasing $J_z$, there is a first order transition between the FM-$xy$ and the FM-$z$ order at low temperature.

A weak but finite Zeeman field $h$ would favor the FM-$z$ over FM-$xy$ order. In Fig. 1 we can see that the FM-$xy$ phase shrinks in the phase diagram under nonzero $h$. With a nonzero $h$, there is no longer a sharp transition between a FM-$z$ order at low temperature, and a disordered phase at high temperature, but one can still define a crossover temperature $T'_c$ which corresponds to $\langle S^z \rangle$ drops to $1/10$. A plot of $T_c$ and $T'_c$ on the $T - J_z$ plane for different values of $h$ is shown in fig.1.

We also show how $\phi_x$ and $\phi_z$ evolve under increasing $h$ (Fig. 2). $J_z$ is chosen to be close to the phase boundary between the FM-$xy$ order and the FM-$z$. One can see that a relatively weak $h$ can completely suppress the FM-$xy$ order, and turn the system into the FM-$z$ order, and the expectation value of $S^z$ rapidly saturates. All these are consistent with what was observed experimentally.

The global phase diagram Eq. 1 may include various much more exotic phases, beyond the semiclassical ferromagnetic phases being discussed here. For example, if the system is not fixed at total filling $\nu_{ex} = 1$, each flavor of the exciton can be viewed as a hard-core boson with large $U$ in Eq. 1, and the system may be viewed as an effective two-orbital spin-1/2 models as $\hat{n}_{i,\alpha} = 1, 0$ can be mapped to spin-up or down of a spin-1/2 degree of freedom. The entire system may also be viewed as a SU(4) spin model with certain anisotropy. SU($N$) quantum magnets have attracted enormous interests since the early days of spin liquids [5–13], and the SU(4) quantum magnets are of particular interests as they can be engineered in transition metal oxides with spin and orbital degrees of freedom [14, 15], cold atom systems [16–18], and graphene based moire systems [2, 19–21]. It was shown numerically that the system may support many possible exotic phases [22–28]. Also, the fluxes seen by the boson hoppingsin Eq. 1 will lead to spin-chirality terms in the effective Hamiltonian of the two-orbital spin-1/2 models, which may lead to topological orders according to various recent numerical works [29–33].

### III. CLASSICAL $O(3)$ MODEL

Since the experiment is performed at finite temperature, we can also treat the spin vector $S^a$ as a classical $O(3)$ vector with unit length. The Hamiltonian is still given by Eq. 4, but now $S_i^a$ are components of a unit $O(3)$ classical vector.

The mean field decomposition proceeds as before and the mean field free energy is given as:

$$\begin{aligned}
Z &= \text{Tr}\{\exp(-\beta H_{MF})\} \\
&= \prod_i \int D\vec{S} \exp\{\beta(6J_\perp \phi_x S^x + 6J_\perp \phi_y S^y + 12 J_z \phi_z S^z + hS^z\} \\
&\quad \times \exp\beta(-\sum_{<ij>} J_\perp \phi_x^2 + J_\perp \phi_y^2 - \sum_{<ij>,\ll ij\gg} J_z \phi_z^2) \\
&= \left(\frac{4\pi}{\beta\sqrt{(6J_\perp \phi_x)^2 + (12J_z \phi_z + h)^2}} \sinh\beta\sqrt{(6J_\perp \phi_x)^2 + (12J_z \phi_z + h)^2} \exp\left(-\beta(3J_\perp \phi_x^2 + 3J_\perp \phi_y^2 + 6J_z \phi_z^2)\right)\right)^N
\end{aligned} \quad (10)$$

We have again used the $O(2)$ symmetry to set $\phi_y = 0$. The free energy density is

$$f = -\frac{1}{\beta}\ln\left(\frac{4\pi}{\beta\sqrt{(6J_\perp \phi_x)^2 + (12J_z \phi_z + h)^2}} \sinh\beta\sqrt{(6J_\perp \phi_x)^2 + (12J_z \phi_z + h)^2}\right) + 3J_\perp \phi_x^2 + 6J_z \phi_z^2 \quad (11)$$

The consistency equations for $\phi_x$ and $\phi_z$ then follow

$$6J_\perp \phi_x \frac{(6J_\perp \phi_x)^2 + (12J_z \phi_z + h)^2}{\sinh\beta\sqrt{(6J_\perp \phi_x)^2 + (12J_z \phi_z + h)^2}}$$
$$\left(\frac{\cosh\beta\sqrt{(6J_\perp \phi_x)^2 + (12J_z \phi_z + h)^2}}{(6J_\perp \phi_x)^2 + (12J_z \phi_z + h)^2} - \frac{\sinh\beta\sqrt{(6J_\perp \phi_x)^2 + (12J_z \phi_z + h)^2}}{\beta((6J_\perp \phi_x)^2 + (12J_z \phi_z + h)^2)^{\frac{3}{2}}}\right) = \phi_x \quad (12)$$

$$(12J_z \phi_z + h)\frac{(6J_\perp \phi_x)^2 + (12J_z \phi_z + h)^2}{\sinh\beta\sqrt{(6J_\perp \phi_x)^2 + (12J_z \phi_z + h)^2}}$$
$$\left(\frac{\cosh\beta\sqrt{(6J_\perp \phi_x)^2 + (12J_z \phi_z + h)^2}}{(6J_\perp \phi_x)^2 + (12J_z \phi_z + h)^2} - \frac{\sinh\beta\sqrt{(6J_\perp \phi_x)^2 + (12J_z \phi_z + h)^2}}{\beta((6J_\perp \phi_x)^2 + (12J_z \phi_z + h)^2)^{\frac{3}{2}}}\right) = \phi_z. \quad (13)$$

The consistency equation can be solved numerically, and the physics is qualitatively the same as the quantum mean field theory presented in the previous section.

---

[1] E. Altman, W. Hofstetter, E. Demler, and M. D. Lukin, Phase diagram of two-component bosons on an optical lattice, New Journal of Physics **5**, 113 (2003).

[2] H. C. Po, L. Zou, A. Vishwanath, and T. Senthil, Origin of mott insulating behavior and superconductivity in twisted bilayer graphene, Phys. Rev. X **8**, 031089 (2018).
[3] Y.-H. Zhang and T. Senthil, Bridging hubbard model physics and quantum hall physics in trilayer graphene/$h-$BN moiré superlattice, Phys. Rev. B **99**, 205150 (2019).
[4] F. Wang, F. Pollmann, and A. Vishwanath, Extended supersolid phase of frustrated hard-core bosons on a triangular lattice, Phys. Rev. Lett. **102**, 017203 (2009).
[5] J. B. Marston and I. Affleck, Large-$n$ limit of the hubbard-heisenberg model, Phys. Rev. B **39**, 11538 (1989).
[6] N. Read and S. Sachdev, Some features of the phase diagram of the square lattice su(n) antiferromagnet, Nuclear Physics B **316**, 609 (1989).
[7] S. Sachdev, Kagome´- and triangular-lattice heisenberg antiferromagnets: Ordering from quantum fluctuations and quantum-disordered ground states with unconfined bosonic spinons, Phys. Rev. B **45**, 12377 (1992).
[8] N. Read and S. Sachdev, Large-n expansion for frustrated quantum antiferromagnets, Phys. Rev. Lett. **66**, 1773 (1991).
[9] N. Read and S. Sachdev, Spin-peierls, valence-bond solid, and néel ground states of low-dimensional quantum antiferromagnets, Phys. Rev. B **42**, 4568 (1990).
[10] N. Read and S. Sachdev, Valence-bond and spin-peierls ground states of low-dimensional quantum antiferromagnets, Phys. Rev. Lett. **62**, 1694 (1989).
[11] D. S. Rokhsar, Quadratic quantum antiferromagnets in the fermionic large-n limit, Phys. Rev. B **42**, 2526 (1990).
[12] M. Hermele, T. Senthil, and M. P. A. Fisher, Algebraic spin liquid as the mother of many competing orders, Phys. Rev. B **72**, 104404 (2005).
[13] C. Xu, Liquids in multiorbital SU(n) magnets made up of ultracold alkaline-earth atoms, Phys. Rev. B **81**, 144431 (2010).
[14] S. K. Pati, R. R. P. Singh, and D. I. Khomskii, Alternating spin and orbital dimerization and spin-gap formation in coupled spin-orbital systems, Phys. Rev. Lett. **81**, 5406 (1998).
[15] K. I. Kugel' and D. I. KhomskiĬ, The jahn-teller effect and magnetism: transition metal compounds, Soviet Physics Uspekhi **25**, 231 (1982).
[16] C. Wu, J.-p. Hu, and S.-c. Zhang, Exact so(5) symmetry in the spin-3/2 fermionic system, Phys. Rev. Lett. **91**, 186402 (2003).
[17] C. Wu, Competing orders in one-dimensional spin-3/2 fermionic systems, Phys. Rev. Lett. **95**, 266404 (2005).
[18] A. V. Gorshkov, M. Hermele, V. Gurarie, C. Xu, P. S. Julienne, J. Ye, P. Zoller, E. Demler, M. D. Lukin, and A. M. Rey, Two-orbital s u(n) magnetism with ultracold alkaline-earth atoms, Nature Physics **6**, 289 (2010).
[19] C. Xu and L. Balents, Topological superconductivity in twisted multilayer graphene, Phys. Rev. Lett. **121**, 087001 (2018).
[20] N. F. Q. Yuan and L. Fu, Model for the metal-insulator transition in graphene superlattices and beyond, Phys. Rev. B **98**, 045103 (2018).
[21] Y.-Z. You and A. Vishwanath, Superconductivity from valley fluctuations and approximate SO(4) symmetry in a weak coupling theory of twisted bilayer graphene, npj Quantum Materials **4**, 10.1038/s41535-019-0153-4 (2019).
[22] Y. Q. Li, M. Ma, D. N. Shi, and F. C. Zhang, Su(4) theory for spin systems with orbital degeneracy, Phys. Rev. Lett. **81**, 3527 (1998).
[23] C. Xu and C. Wu, Resonating plaquette phases in su(4) heisenberg antiferromagnet, Phys. Rev. B **77**, 134449 (2008).
[24] P. Corboz, M. Lajkó, A. M. Läuchli, K. Penc, and F. Mila, Spin-orbital quantum liquid on the honeycomb lattice, Phys. Rev. X **2**, 041013 (2012).
[25] K. Penc, M. Mambrini, P. Fazekas, and F. Mila, Quantum phase transition in the su(4) spin-orbital model on the triangular lattice, Phys. Rev. B **68**, 012408 (2003).
[26] M. Hermele, V. Gurarie, and A. M. Rey, Mott insulators of ultracold fermionic alkaline earth atoms: Underconstrained magnetism and chiral spin liquid, Phys. Rev. Lett. **103**, 135301 (2009).
[27] A. Keselman, B. Bauer, C. Xu, and C.-M. Jian, Emergent fermi surface in a triangular-lattice su(4) quantum antiferromagnet, Phys. Rev. Lett. **125**, 117202 (2020).
[28] Y.-H. Zhang, D. N. Sheng, and A. Vishwanath, Su(4) chiral spin liquid, exciton supersolid, and electric detection in moiré bilayers, Phys. Rev. Lett. **127**, 247701 (2021).
[29] M. Greiter and R. Thomale, Non-abelian statistics in a quantum antiferromagnet, Phys. Rev. Lett. **102**, 207203 (2009).
[30] B. Bauer, L. Cincio, B. Keller, M. Dolfi, G. Vidal, S. Trebst, and A. Ludwig, Chiral spin liquid and emergent anyons in a kagome lattice mott insulator, Nature Communications **5**, 10.1038/ncomms6137 (2014).
[31] J.-Y. Chen, L. Vanderstraeten, S. Capponi, and D. Poilblanc, Non-abelian chiral spin liquid in a quantum antiferromagnet revealed by an ipeps study, Phys. Rev. B **98**, 184409 (2018).
[32] Z.-X. Liu, H.-H. Tu, Y.-H. Wu, R.-Q. He, X.-J. Liu, Y. Zhou, and T.-K. Ng, Non-abelian $s=1$ chiral spin liquid on the kagome lattice, Phys. Rev. B **97**, 195158 (2018).
[33] W.-W. Luo, Y. Huang, D. N. Sheng, and W. Zhu, Global quantum phase diagram and non-abelian chiral spin liquid in a spin-3/2 square lattice antiferromagnet (2022).